\documentclass{article}%
\usepackage{graphicx}
\usepackage{amsmath}
\usepackage{amsfonts}
\usepackage{amssymb}%
\providecommand{\U}[1]{\protect\rule{.1in}{.1in}}

\begin{document}

\author{Antony Valentini\\Augustus College}

\begin{center}
{\LARGE Beyond the Born rule in quantum gravity}

\bigskip

\bigskip

\bigskip

Antony Valentini\footnote{antonyv@clemson.edu}

\textit{Augustus College,}

\textit{14 Augustus Road, London SW19 6LN, UK.}

\textit{Department of Physics and Astronomy,}

\textit{Clemson University, Kinard Laboratory,}

\textit{Clemson, SC 29634-0978, USA.}

\bigskip

\bigskip
\end{center}

\bigskip\bigskip

\bigskip

\bigskip

We have recently developed a new understanding of probability in quantum
gravity. In this paper we provide an overview of this new approach and its
implications. Adopting the de Broglie-Bohm pilot-wave formulation of quantum
physics, we argue that there is no Born rule at the fundamental level of
quantum gravity with a non-normalisable Wheeler-DeWitt wave functional $\Psi$.
Instead the universe is in a perpetual state of quantum nonequilibrium with a
probability density $P\neq\left\vert \Psi\right\vert ^{2}$. Dynamical
relaxation to the Born rule can occur only after the early universe has
emerged into a semiclassical or Schr\"{o}dinger approximation, with a
time-dependent and normalisable wave functional $\psi$, for non-gravitational
systems on a classical spacetime background. In that regime the probability
density $\rho$ can relax towards $\left\vert \psi\right\vert ^{2}$ (on a
coarse-grained level). Thus the pilot-wave theory of gravitation supports the
hypothesis of primordial quantum nonequilibrium, with relaxation to the Born
rule taking place soon after the big bang. We also show that
quantum-gravitational corrections to the Schr\"{o}dinger approximation allow
quantum nonequilibrium $\rho\neq\left\vert \psi\right\vert ^{2}$ to be created
from a prior equilibrium ($\rho=\left\vert \psi\right\vert ^{2}$) state. Such
effects are very tiny and difficult to observe in practice.

\bigskip

\bigskip

\bigskip

\bigskip

\bigskip

Published in special issue \textit{Pilot-wave and beyond: Louis de Broglie and
David Bohm's quest for a quantum ontology}, ed. A. Drezet, Found. Phys.
\textbf{53}, 6 (2023).

\bigskip

\bigskip

\section{Introduction}

It has long been known that the pilot-wave theory of de Broglie and Bohm
provides us with an objective and deterministic account of quantum physics
\cite{deB28,BV09,B52,Holl93,AVORE}. Historically, the theory was constructed
by de Broglie in a series of papers from 1922 to 1927, culminating in a
pilot-wave dynamics for a many-body system.\footnote{By de Broglie's own
account his ideas originated in a paper of 1922 on blackbody radiation,
although his first paper on pilot-wave theory proper did not appear until
1923, culminating in his theory of a many-body system presented at the 1927
Solvay conference (see ref. \cite{BV09}, chapter 2).} The theory was revived
by Bohm in 1952, who extended the dynamics to field theory and, crucially,
showed in detail how the theory accounts for the general quantum theory of
measurement \cite{B52}. Despite this success, there remains a long-standing
controversy concerning the status of the Born probability rule in this theory.
In recent work we have argued that the status of the Born rule in pilot-wave
theory changes radically when we consider a regime in which
quantum-gravitational effects are important \cite{AV21}. In this paper we
provide an overview of these new ideas and results, with a minimum of
technicalities, and with an emphasis on the conceptual implications.

In pilot-wave theory a system with configuration-space wave function
$\psi(q,t)$ has an actual trajectory $q(t)$ whose velocity $v=dq/dt$ is
determined by de Broglie's law of motion or `guidance equation', where for
systems with conventional Hamiltonians $v$ is proportional to the gradient
$\partial_{q}S$ of the phase $S$ of $\psi=\left\vert \psi\right\vert e^{iS}$.
For an ensemble of systems with the same wave function $\psi(q,t)$, the
ensemble distribution $\rho(q,t)$ of configurations is usually assumed to be
given by the Born rule%
\begin{equation}
\rho=\left\vert \psi\right\vert ^{2}\ . \label{Born_0}%
\end{equation}
It is a simple consequence of the equations of motion that if (\ref{Born_0})
holds at an initial time $t=0$ it will hold for all $t$. On these grounds in
the 1920s de Broglie simply took (\ref{Born_0}) as an assumption with no
further explanation or justification \cite{BV09}. This stance was however
criticised by Pauli and by Keller, in 1953, who argued that such an initial
condition was unjustified in a deterministic theory and should be derived from
the dynamics \cite{Pauli53,Keller53}. This criticism was partially met by Bohm
in the same year when he argued that an ensemble of two-level molecules would
relax to the state (\ref{Born_0}) when subjected to random collisions
\cite{B53}. However, no general argument for relaxation was given. In 1954,
citing difficulties with understanding relaxation to the Born rule, Bohm and
Vigier abandoned the original deterministic theory and introduced random
(subquantum) `fluid fluctuations' that drive relaxation to the Born rule for a
general system \cite{BV54}. Since then, most authors have simply adopted de
Broglie's original position, with (\ref{Born_0}) in effect taken as an
additional postulate (alongside the Schr\"{o}dinger equation for $\psi$ and de
Broglie's law for $v$) \cite{Holl93,DGZ92,DT09,Tum21}.

This author has long argued that simply postulating (\ref{Born_0}) is a
mistake, akin to artificially restricting classical mechanics to a state of
thermal equilibrium
\cite{AV91a,AV91b,AV92,AV96,AV01,AV02a,AV02b,VW05,PV06,AV09,Allori20}. In
pilot-wave theory the Born rule (\ref{Born_0}) really describes a state of
statistical equilibrium, or `quantum equilibrium', analogous for example to
the Maxwell distribution of molecular speeds for a gas in thermal equilibrium.
Just as a classical ensemble can be in thermal nonequilibrium, with a
distribution of velocities different from that of Maxwell, in pilot-wave
theory an ensemble can be in quantum nonequilibrium with a distribution of
configurations%
\begin{equation}
\rho\neq\left\vert \psi\right\vert ^{2} \label{noneq_0}%
\end{equation}
different from that of Born. For such an ensemble the statistical predictions
of textbook quantum mechanics would fail -- raising the question of why such
nonequilibrium phenomena have never been observed in the laboratory. The
answer, at least as proposed by this author in 1991, is that all the systems
we have access to have a long and violent history that traces back ultimately
to the big bang. During that time there has been ample opportunity for
dynamical relaxation $\rho\rightarrow\left\vert \psi\right\vert ^{2}$ to take
place (on a coarse-grained level) -- a process of `quantum relaxation' that is
broadly analogous to classical thermal relaxation, and which presumably
occurred in the early universe. This process has been studied in general terms
and has been observed to take place efficiently in a wide range of numerical
simulations
\cite{AV91a,AV91b,AV92,AV01,VW05,Allori20,EC06,TRV12,SC12,ACV14,ECT17,Drezet21,Lust21,Lust22}%
. We can then understand why the Born rule holds to high accuracy today. At
the same time, we understand that pilot-wave theory also contains a wider
nonequilibrium physics, which may have been active in the early universe, and
which could have left discernible traces today -- in the form of anomalies in
the cosmic microwave background (CMB), as well as in relic cosmological
particles that might today still display violations of the Born rule
\cite{AV01,AV07,AV08a,AV10,CV13,CV15,AV15,CV16,UV15,UV20}.

On this view, quantum physics is merely a special case of a much wider physics
in which the Born rule is broken. That wider physics allows violations of the
uncertainty principle as well as practical nonlocal signalling
\cite{AV91b,AV02a,AV02b,PV06}. According to pilot-wave theory, at least when
correctly interpreted, textbook quantum mechanics is merely an effective
theory that emerges in the state of quantum equilibrium. Many supposedly
fundamental quantum constraints are really peculiarities of equilibrium and
are broken for more general ensembles.

An alternative view has, however, long been championed by the `Bohmian
mechanics' school of de Broglie-Bohm theory -- a distinctive approach to the
subject first proposed by D\"{u}rr, Goldstein and Zangh\`{\i} in 1992
\cite{DGZ92,DT09,Tum21}. In this approach, the wave function $\Psi$ of the
whole universe is used to define a (supposedly) fundamental probability (or
`typicality')\ measure $\left\vert \Psi\right\vert ^{2}$, from which one can
readily derive the Born rule (\ref{Born_0}) for subsystems with an effective
wave function $\psi$. On this view the Born rule is built into the theory and
there is no prospect of ever finding nonequilibrium violations, not even in
the early universe. While this approach has been influential among some
philosophers \cite{Albert15,Gold21}, the argument is essentially circular: the
Born rule is derived for subsystems only by assuming the Born rule for the
whole universe at the initial time $t=0$. There is no reason why our universe
should have started with those particular initial conditions -- whether or not
the universe began in equilibrium or nonequilibrium is ultimately an empirical
question to be decided by observation and experiment, not by theoretical or
philosophical fiat \cite{AV96,AV01,Allori20}.

As we will see in this paper, the controversy over the Born rule in pilot-wave
theory changes drastically when we consider a regime where quantum gravity is
important. For in that regime there simply is no normalisable physical
probability (or typicality) measure $\left\vert \Psi\right\vert ^{2}$ for the
whole universe, and the (circular) argument employed by the Bohmian mechanics
school can no longer even be formulated. Instead, normalisable wave functions
$\psi$ emerge only in the semiclassical regime -- for systems evolving on a
classical spacetime background -- and in that regime the Born rule
(\ref{Born_0}) can emerge by a dynamical process of quantum relaxation
\cite{AV21}. In this way, considerations from quantum gravity vindicate the
hypothesis of quantum nonequilibrium at the big bang, with relaxation to the
Born rule taking place only afterwards.

Before presenting technical details of how all this works, let us first sketch
the key ideas in simple terms. In canonical quantum gravity the geometry of
3-space is described by a metric tensor $g_{ij}$.\footnote{In this paper we
employ the traditional metric representation of the gravitational field. We
expect similar conclusions to hold in loop quantum gravity \cite{Rov04}.} If
we include a matter field $\phi$ we might expect the system to have a wave
function (or functional) $\Psi\lbrack g_{ij},\phi,t]$ obeying a time-dependent
Schr\"{o}dinger equation $i\partial\Psi/\partial t=\hat{H}\Psi$ (with an
appropriate Hamiltonian $\hat{H}$ and time parameter $t$). Instead, when we
quantise the gravitational field we obtain a wave functional $\Psi\lbrack
g_{ij},\phi]$ obeying a time-independent Wheeler-DeWitt equation
\cite{DeW67,Kief12}%
\begin{equation}
\mathcal{\hat{H}}\Psi=0 \label{W-D}%
\end{equation}
(with an appropriate Hamiltonian density operator $\mathcal{\hat{H}}%
$).\footnote{As we will see there is also a constraint on $\Psi$ guaranteeing
coordinate invariance.} Time makes no appearance in the equations. After more
than half a century since it was first written down, the physical
interpretation of this `timeless' theory remains controversial.

Most workers in the field agree that a time-dependent Schr\"{o}dinger equation
$i\partial\psi/\partial t=\hat{H}\psi$ for a conventional wave functional
$\psi\lbrack\phi,t]$ can emerge only in a semiclassical regime for a quantum
field $\phi$ propagating on a classical background spacetime. There is,
however, controversy over precisely how an effective time parameter $t$ can
emerge from a fundamentally timeless theory. This question is known in the
quantum gravity literature as the `problem of time'
\cite{UW89,Ish91,Kuch92,Ish93,Kuch99,Anders17,KP22}.

Quantum-gravitational effects are expected to be significant at sufficiently
early times in our cosmological history (certainly within a Planck time
$t_{\mathrm{P}}\sim10^{-43}\ \sec$ after the beginning). In such a deep
quantum-gravity regime, the Wheeler-DeWitt equation (\ref{W-D}) must be
applied. Soon afterwards we expect the universe to emerge into a semiclassical
or `Schr\"{o}dinger' regime, in which a conventional time-dependent wave
equation can be applied. Previous discussion of the Born rule and of quantum
relaxation in pilot-wave theory has taken place within the semiclassical or
Schr\"{o}dinger approximation. Outside that approximation, however, the
discussion must be carefully revised.

A pilot-wave theory of quantum gravity can be written down by supplementing
the Wheeler-DeWitt equation (\ref{W-D}) with de Broglie-Bohm trajectories
$g_{ij}(t)$ whose velocity $\partial g_{ij}/\partial t$ is proportional to a
generalised phase gradient (see Section 5)
\cite{Vink92,Hor94,Sht96,PN05,PNF13}.\footnote{For a recent review see ref.
\cite{PN21}.} We might then expect $\left\vert \Psi\lbrack g_{ij}%
,\phi]\right\vert ^{2}$ to define an equilibrium Born-rule probability density
\cite{Vink92,DS20}. But, as we will see, this cannot be correct. The
mathematical structure of (\ref{W-D}) ensures that the density $\left\vert
\Psi\lbrack g_{ij},\phi]\right\vert ^{2}$ cannot be normalised and so cannot
be a physical probability distribution. This point has caused controversy and
confusion in the literature. In our view, from a pilot-wave perspective, the
implication is clear: in quantum gravity there simply is no physical Born-rule
equilibrium state \cite{AV21,AV14}. A physical probability density
$P[g_{ij},\phi,t]$ (for a theoretical ensemble) must be normalisable by
definition. Therefore it must differ from $\left\vert \Psi\lbrack g_{ij}%
,\phi]\right\vert ^{2}$ at all times. We may say that the deep quantum-gravity
regime is in a perpetual state of quantum nonequilibrium%
\begin{equation}
P\neq\left\vert \Psi\right\vert ^{2}\ .
\end{equation}
In this regime there can be no quantum relaxation and no state of quantum equilibrium.

As we will see there are two immediate implications. First, as the early
universe emerges from the deep quantum-gravity regime and settles into a
semiclassical regime described by the Schr\"{o}dinger approximation, we can
expect fields $\phi$ propagating on the classical background to be in a state
of quantum nonequilibrium $\rho\lbrack\phi,t]\neq\left\vert \psi\lbrack
\phi,t]\right\vert ^{2}$ -- where $\psi$ is the effective (normalisable)
Schr\"{o}dinger wave functional for $\phi$ and the probability $\rho$ emerges
from $P$ as a conditional probability. Second, quantum relaxation as
previously understood can begin to take place only after the universe has
settled into a conventional Schr\"{o}dinger regime. Thus, even though there is
no fundamental Born-rule equilibrium state in quantum gravity, we still
recover quantum relaxation in the Schr\"{o}dinger approximation -- and so we
can still explain the Born rule as we see it today.

There is another remarkable result of this analysis. It is well known that the
emergent Schr\"{o}dinger equation is subject to small quantum-gravitational
corrections appearing in the effective Hamiltonian $\hat{H}$. Perhaps
surprisingly, some of the correction terms are non-Hermitian
\cite{KS91,KK12,Bini13,BKM16}. Such terms are of course inconsistent with
standard quantum mechanics, since the norm of $\psi$ is no longer conserved
and $\left\vert \psi\right\vert ^{2}$ cannot be interpreted as a probability
density in the usual way. For this reason, in previous studies such terms have
been dropped, with no clear justification. In pilot-wave theory, in contrast,
there is no inconsistency: such terms simply generate a gravitational
instability of the Born rule, whereby an initial density $\rho=\left\vert
\psi\right\vert ^{2}$ can evolve into a final density $\rho\neq\left\vert
\psi\right\vert ^{2}$. As we will see, such effects are extremely small, but
observable at least in principle. This means that, when quantum gravity is
taken into account, it is no longer the case that once quantum equilibrium is
reached we are trapped in that state forever. There is a way out, at least in principle.

To summarise, in this paper we have three new ideas to present:

\begin{enumerate}
\item In quantum gravity there is no Born rule for a timeless Wheeler-DeWitt
wave functional $\Psi$ and the system is in a perpetual state of quantum
nonequilibrium $P\neq\left\vert \Psi\right\vert ^{2}$.

\item The Born rule $\rho=\left\vert \psi\right\vert ^{2}$ can emerge by
quantum relaxation only in a semiclassical or Schr\"{o}dinger approximation,
for systems with an effective time-dependent wave function $\psi$ on a
classical background spacetime.

\item Tiny quantum-gravitational corrections to the Schr\"{o}dinger
approximation can make the Born rule unstable, with initial distributions
$\rho=\left\vert \psi\right\vert ^{2}$ evolving to final distributions
$\rho\neq\left\vert \psi\right\vert ^{2}$.
\end{enumerate}

To develop the details, we begin with a brief outline of some essential formalism.

\section{Quantum gravity and quantum cosmology}

In this section we provide a brief summary of the essential formalism of
canonical quantum gravity \cite{DeW67,Kief12}, together with a simple model of
quantum cosmology.

\subsection{Canonical quantum gravity}

The canonical quantisation of the gravitational field begins with a `3+1'
foliation of classical spacetime by spacelike slices $\Sigma(t)$ labelled by a
time parameter $t$. This can always be done (generally nonuniquely) for a
spacetime that is `globally-hyperbolic'. The line element then takes the form%
\begin{equation}
d\tau^{2}=(N^{2}-N_{i}N^{i})dt^{2}-2N_{i}dx^{i}dt-g_{ij}dx^{i}dx^{j}\ ,
\label{ADM}%
\end{equation}
where $N$, $N^{i}$ are respectively the `lapse function' and `shift vector',
while $g_{ij}$ is the 3-metric on $\Sigma(t)$. For simplicity we can take
$N^{i}=0$ -- so that lines of constant $x^{i}$ are normal to the slices --
provided such lines do not encounter singularities. The object to be quantised
is then a spatial 3-geometry represented by $g_{ij}$.

Beginning with the usual Einstein-Hilbert action, standard quantisation
methods lead to the Wheeler-DeWitt equation, which for the pure gravitational
field reads\footnote{For a `functional' $\Psi\lbrack\phi]$ (mapping a function
$\phi(x)$ to a complex number $\Psi$) the functional derivative $\delta
\Psi/\delta\phi(x)$ at a spatial point $x$ is defined by $\delta\Psi=\int
d^{3}x\ \left[  \delta\Psi/\delta\phi(x)\right]  \delta\phi(x)$ for arbitrary
infinitesimal variations $\delta\phi(x)$.}%
\begin{equation}
\left(  -G_{ijkl}\frac{\delta^{2}}{\delta g_{ij}\delta g_{kl}}-\sqrt
{g}R\right)  \Psi=0\ , \label{WD1}%
\end{equation}
where $\Psi=\Psi\lbrack g_{ij}]$ and%
\begin{equation}
G_{ijkl}={\frac{1}{2}}g^{-1/2}(g_{ik}g_{jl}+g_{il}g_{jk}-g_{ij}g_{kl})\ .
\end{equation}
We have written the kinetic term with a specific operator ordering but it
should be understood that the ordering is ambiguous. The wave functional is
also subject to a constraint%
\begin{equation}
-2D_{j}\frac{\delta\Psi}{\delta g_{ij}}=0 \label{cons1}%
\end{equation}
associated with spatial coordinate invariance (where $D_{j}$ is a spatial
covariant derivative). This constraint ensures that $\Psi$ is a function of
the coordinate-independent 3-geometry and not of the coordinate-dependent
3-metric. Writing (\ref{WD1}) and (\ref{cons1}) as%
\[
\mathcal{\hat{H}}\Psi=0\ ,\ \ \ \ \ \mathcal{\hat{H}}^{i}\Psi=0\ ,
\]
the total Hamiltonian operator is given by%
\begin{equation}
\hat{H}=\int d^{3}x\ (N\mathcal{\hat{H}}+N_{i}\mathcal{\hat{H}}^{i})\ .
\end{equation}

In the presence of a scalar matter field $\phi$ with potential $\mathcal{V}%
(\phi)$ we have an extended Wheeler-DeWitt equation%
\begin{equation}
(\mathcal{\hat{H}}_{g}+\mathcal{\hat{H}}_{\phi})\Psi=0 \label{W-D_ext}%
\end{equation}
for the wave functional $\Psi=\Psi\lbrack g_{ij},\phi]$, where the
gravitational term%
\begin{equation}
\mathcal{\hat{H}}_{g}=-G_{ijkl}\frac{\delta^{2}}{\delta g_{ij}\delta g_{kl}%
}-\sqrt{g}R
\end{equation}
in $\mathcal{\hat{H}}=\mathcal{\hat{H}}_{g}+\mathcal{\hat{H}}_{\phi}$ is
supplemented by a matter term%
\begin{equation}
\mathcal{\hat{H}}_{\phi}=\frac{1}{2}\sqrt{g}\left(  -\frac{1}{g}\frac
{\delta^{2}}{\delta\phi^{2}}+g^{ij}\partial_{i}\phi\partial_{j}\phi\right)
+\sqrt{g}\mathcal{V}\ , \label{matter_H}%
\end{equation}
while the constraint $\mathcal{\hat{H}}^{i}\Psi=0$ corresponding to
(\ref{cons1}) takes the form%
\begin{equation}
-2D_{j}\frac{\delta\Psi}{\delta g_{ij}}+\partial^{i}\phi\frac{\delta\Psi
}{\delta\phi}=0\ . \label{cons2}%
\end{equation}

\subsection{A simple model of quantum cosmology}

It will be helpful to illustrate our ideas with a simple model of quantum cosmology.

Consider an expanding flat and homogeneous universe with scale factor $a(t)$
and spacetime line element\footnote{Here $dx^{2}=(dx^{1})^{2}+(dx^{2}%
)^{2}+(dx^{3})^{2}$.}%

\begin{equation}
d\tau^{2}=dt^{2}-a^{2}dx^{2}\ . \label{Fried}%
\end{equation}
We assume that the universe contains a homogeneous matter field $\phi$ with a
potential $\mathcal{V}(\phi)$. We then have a `mini-superspace' model with two
degrees of freedom $(a,\phi)$.

This system has a Lagrangian \cite{BKM16,Vent14}%
\begin{equation}
L=-\frac{1}{2}m_{\mathrm{P}}^{2}a\dot{a}^{2}+\frac{1}{2}a^{3}\dot{\phi}%
^{2}-a^{3}\mathcal{V}\ ,
\end{equation}
where $m_{\mathrm{P}}^{2}=3/4\pi G$ is the square of a (rescaled) Planck mass.
This implies canonical momenta%
\begin{equation}
p_{a}=-m_{\mathrm{P}}^{2}a\dot{a}\ ,\ \ \ \ \ p_{\phi}=a^{3}\dot{\phi}
\label{can_moma_qucosmo}%
\end{equation}
and a Hamiltonian%
\begin{equation}
H=-\frac{1}{2m_{\mathrm{P}}^{2}}\frac{1}{a}p_{a}^{2}+\frac{1}{2a^{3}}p_{\phi
}^{2}+a^{3}\mathcal{V}%
\end{equation}
(noting the sign difference between the kinetic terms).

Promoting the canonical momenta to operators $\hat{p}_{a}=-i\partial/\partial
a$ and $\hat{p}_{\phi}=-i\partial/\partial\phi$, and choosing the factor
ordering $\frac{1}{a^{2}}\hat{p}_{a}a\hat{p}_{a}$, the Wheeler-DeWitt equation
$\hat{H}\Psi=0$ for $\Psi(a,\phi)$ reads \cite{BKM16}%

\begin{equation}
\frac{1}{2m_{\mathrm{P}}^{2}}\frac{1}{a}\frac{\partial}{\partial a}\left(
a\frac{\partial\Psi}{\partial a}\right)  -\frac{1}{2a^{2}}\frac{\partial
^{2}\Psi}{\partial\phi^{2}}+a^{4}\mathcal{V}\Psi=0\ . \label{WD_mini}%
\end{equation}

\section{Difficulties with the Born rule in quantum gravity}

In this section we discuss some of the key difficulties with trying to
interpret $\left\vert \Psi\right\vert ^{2}$ as a probability density for a
Wheeler-DeWitt wave funtional $\Psi$.

\subsection{Why $\left\vert \Psi\right\vert ^{2}$ is non-normalisable}

We have said that, for solutions $\Psi$ of the Wheeler-DeWitt equation, the
quantity $\left\vert \Psi\right\vert ^{2}$ cannot be a physical probability
density because it is non-normalisable. To see why $\left\vert \Psi\right\vert
^{2}$ cannot be normalised, consider for simplicity the case of pure
gravitation. The Wheeler-DeWitt equation (\ref{WD1}) for the wave functional
$\Psi\lbrack g_{ij}]$ on the space of 3-metrics $g_{ij}$ is mathematically
analogous to the single-particle Klein-Gordon equation%
\begin{equation}
-\frac{\partial^{2}\psi}{\partial t^{2}}+\delta^{ij}\frac{\partial^{2}\psi
}{\partial x^{i}\partial x^{j}}-m^{2}\psi=0 \label{K-G}%
\end{equation}
for a wave $\psi(x,t)$ on Minkowski spacetime. The analogy can be traced to
the indefinite character of the `DeWitt metric' $G_{ijkl}$ \cite{DeW67}. This
means that (\ref{WD1}) is formally analogous to an infinite-dimensional
Klein-Gordon equation with a `mass-squared' term $g^{1/2}R$. As a result, the
integral $\int Dg\ \left\vert \Psi\lbrack g_{ij}]\right\vert ^{2}$ over the
whole space of 3-metrics necessarily diverges, just as the integral $\int
d^{3}x\int dt\ \left\vert \psi(x,t)\right\vert ^{2}$ over the whole of
spacetime necessarily diverges. It might be thought that the divergence could
be removed by an appropriate regularisation. But the divergence is deeper than
that, reflecting a basic fact about wave propagation. Solutions $\psi(x,t)$ of
the wave equation (\ref{K-G}) can be localised with respect to $x$ but not
with respect to $t$, and mathematically the same phenomenon occurs for
solutions $\Psi\lbrack g_{ij}]$ of the wave equation (\ref{WD1}).

What we have just said is slightly simplified. We have not mentioned the
constraint (\ref{cons1}), which ensures that $\Psi\lbrack g_{ij}]$ is not
really a function on the space of coordinate-dependent 3-metrics $g_{ij}$ but
in fact a function on the space of coordinate-independent 3-geometries (a
space commonly referred to as `superspace'). It might then be thought that the
non-normalisability of $\left\vert \Psi\lbrack g_{ij}]\right\vert ^{2}$ could
just be an artifact of having to integrate over an infinite `gauge volume' of
3-metrics representing the same 3-geometry. But in fact the result still
diverges even if we perform a physical integral over the space of 3-geometries
(perhaps by factoring out the gauge volume in some way).

The simplest way to see this is to consider our mini-superspace model of
quantum cosmology (Section 2.2). Each spacelike slice of constant $t$ has a
simple coordinate-independent representation as a flat Euclidean 3-space with
scale factor $a(t)$. If we rewrite the Wheeler-DeWitt equation (\ref{WD_mini})
for $\Psi(a,\phi)$ in terms of $\alpha=\ln a$ we find%
\begin{equation}
\frac{1}{m_{\mathrm{P}}^{2}}\frac{\partial^{2}\Psi}{\partial\alpha^{2}}%
-\frac{\partial^{2}\Psi}{\partial\phi^{2}}+2e^{6\alpha}\mathcal{V}\Psi=0\ .
\label{WD_mini_alpha}%
\end{equation}
This is a two-dimensional Klein-Gordon equation with a potential term. The
free part (ignoring the potential) has the general solution%
\begin{equation}
\Psi_{\mathrm{free}}=f(\phi-m_{\mathrm{P}}\alpha)+g(\phi+m_{\mathrm{P}}%
\alpha)\ , \label{Psi_free}%
\end{equation}
where $f$ and $g$ are packets travelling with the `wave speed'
$c=m_{\mathrm{P}}$ in the two-dimensional `spacetime' $(\alpha,\phi)$. Thus%
\begin{equation}
\int\int d\alpha d\phi\ \left\vert \Psi_{\mathrm{free}}\right\vert ^{2}%
=\infty\ ,
\end{equation}
just as for a Klein-Gordon solution $\psi(x,t)$ we have%
\begin{equation}
\int d^{3}x\int dt\ \left\vert \psi(x,t)\right\vert ^{2}=\infty\ .
\end{equation}
Clearly the non-normalisability of $\Psi$ has nothing to do with the
(technically delicate) issue of unphysical coordinate degrees of freedom. It
is simply a consequence of the Klein-Gordon-like character of the
Wheeler-DeWitt equation and the resulting wave-like propagation in the
mini-superspace $(\alpha,\phi)$. Similar conclusions must hold in the full theory.

\subsection{Naive Schr\"{o}dinger interpretation. I}

It is in fact well known in quantum-gravity circles that $\left\vert
\Psi\right\vert ^{2}$ cannot be a physical probability density
\cite{UW89,Ish91,Kuch92,Ish93,Kuch99,Anders17}. While interpreted as such by
Hawking and collaborators in the 1980s \cite{Hawketal}, this came to be known
as the `naive Schr\"{o}dinger interpretation' \cite{UW89}. Even leaving aside
the question of non-normalisability, the interpretation is problematic because
the putative probability density $\left\vert \Psi\right\vert ^{2}$ is
time-independent. Attempts were made to repair this by some form of
conditioning on a subset of the degrees of freedom, but this `conditional
probability interpretation' led to other problems \cite{Kuch92,Ish93}. In any
case, in our view the interpretation fails from the outset simply because a
non-normalisable density cannot represent a physical probability distribution.

Historically, however, it has been more common to cite another reason for the
failure of the naive interpretation of $\left\vert \Psi\right\vert ^{2}$. It
is claimed that treating $\left\vert \Psi\right\vert ^{2}$ as a conventional
probability is incorrect because `time' is in effect hidden in the metric
degrees of freedom $g_{ij}$. For example, in quantum cosmology with a wave
function $\Psi(a,\phi)$, it is commonplace to treat the scale factor $a$ as an
effective time parameter. We can then try to recover a Born-rule-type
probability for the remaining degrees of freedom at a given value of `time'.
This approach has a long history, dating back to the pioneering work of DeWitt
and Wheeler in the 1960s \cite{DeW67,Wheel68}, but to this day it remains
controversial. Some authors have raised concerns about the \textit{bona fide}
temporal properties of gravitational degrees of freedom \cite{UW89}. For
example, if the scale factor $a$ plays the role of time, what happens to time
in a universe that expands and recontracts? On the other hand, some supporters
of quantum gravity argue that at the deepest level physics is genuinely
timeless, and that our common-sense notions of `time' emerge only
approximately and in certain conditions
\cite{DeW67,Rov90,Rov91,Rov09,Barb94,Hall}. It is however not entirely clear
whether quantum mechanics can be properly applied in a fundamentally timeless
theory \cite{Rov04,Rov91,Hall,Mondra}. A relatively recent and exhaustive
review of the `problem of time' in quantum gravity runs to nearly a thousand
pages and draws no definite conclusions \cite{Anders17}, suggesting that the
problem has yet to be satisfactorily resolved (though some may disagree).

In this paper we offer a new explanation for the failure of the naive
Schr\"{o}dinger interpretation. Our explanation is that, in the deep
quantum-gravity regime, there is no such thing as the Born rule. As we shall
see, we can discuss (time-dependent) probability densities such as
$P[g_{ij},\phi,t]$, but these are not tied to the Born rule and can never be.
Necessarily, $P[g_{ij},\phi,t]\neq\left\vert \Psi\lbrack g_{ij},\phi
]\right\vert ^{2}$ always, since the left-hand side is normalisable (by
definition) and the right-hand side is not. We may say that, at the Planck
scale, a quantum-gravitational universe is in a perpetual state of quantum
nonequilibrium. This, in our view, is the true physical significance of the
non-normalisability of the Wheeler-DeWitt wave functional $\Psi$.

To make sense of this idea, however, we need to look more closely at
pilot-wave theory.

\section{Pilot-wave theory and the Born rule}

When interpreted correctly, pilot-wave theory shows us that the Born rule is
not an axiom or law, but instead represents a statistical state of quantum
equilibrium (analogous to classical thermal equilibrium)
\cite{AV91a,AV91b,AV92,AV96,AV01,AV02a,AV02b,VW05,PV06,AV09,Allori20}. At
least, that is the case in non-gravitational physics. As we shall see, in
pilot-wave gravitation, in contrast, there is no state of quantum equilibrium
and no possibility of obtaining the Born rule, except in the semiclassical regime.

Let us first consider pilot-wave theory for a general system with
configuration $q$ and wave function $\psi(q,t)$ on a background classical
spacetime with global time parameter $t$ (corresponding to a foliation by
spacelike slices $\Sigma(t)$). Here $q$ could represent particle or field
configurations on the spacelike slice at time $t$. The wave function obeys a
time-dependent Schr\"{o}dinger equation%
\begin{equation}
i\frac{\partial\psi}{\partial t}=\hat{H}\psi\label{Sch_gen}%
\end{equation}
with some Hamiltonian operator $\hat{H}$. This implies a continuity equation
for $\left\vert \psi\right\vert ^{2}$,%
\begin{equation}
\frac{\partial\left\vert \psi\right\vert ^{2}}{\partial t}+\partial_{q}\cdot
j=0\ , \label{Cont_psi2_gen}%
\end{equation}
where $\partial_{q}$ is a gradient on configuration space and $j$ satisfies%
\begin{equation}
\partial_{q}\cdot j=2\operatorname{Re}\left(  i\psi^{\ast}\hat{H}\psi\right)
\ . \label{divj}%
\end{equation}
The `current' $j$ can be written in terms of $\psi$ and its functional form
depends on $\hat{H}$ \cite{SV09}.\footnote{The continuity equation
(\ref{Cont_psi2_gen}) can also be derived from Noether's theorem as the local
conservation law associated with a global phase symmetry $\psi\rightarrow\psi
e^{i\theta}$ on configuration space \cite{SV09}.} Given the expression
$j=j\left[  \psi\right]  =j(q,t)$ we can define a velocity field%
\begin{equation}
v(q,t)=\frac{j(q,t)}{|\psi(q,t)|^{2}} \label{deB_vel_gen}%
\end{equation}
and write down a de Broglie guidance equation%
\begin{equation}
\frac{dq}{dt}=v(q,t) \label{deB_gen}%
\end{equation}
for the trajectory $q(t)$. Equations (\ref{Sch_gen}), (\ref{deB_vel_gen}) and
(\ref{deB_gen}) define a deterministic dynamics for a general system with wave
function $\psi(q,t)$.\footnote{We are assuming the wave function has a single
component $\psi$. The method can be readily extended to spin systems with
multi-component wave functions \cite{AVOUP}.} Note that $\psi$ is regarded as
a physical field (or `pilot wave') on configuration space that guides the
trajectory $q(t)$ of a single system. At the fundamental dynamical level there
is no such thing as probability (as in classical mechanics).

If $\hat{H}$ happens to be quadratic in the momenta, we find that $v$ is
proportional to a phase gradient. For example, for a single low-energy
particle of mass $m$ we find%
\begin{equation}
\mathbf{v}=\frac{1}{m}\operatorname{Im}\frac{\mathbf{\nabla}\psi}{\psi}%
=\frac{1}{m}\mathbf{\nabla}S\ ,
\end{equation}
where $\psi=\left\vert \psi\right\vert e^{iS}$, while for a many-body system
the $n$th particle has velocity $\mathbf{v}_{n}=(1/m_{n})\mathbf{\nabla}_{n}S$.

We can now consider an ensemble of systems with the same wave function $\psi$.
The systems evolve according to the velocity field $v$. By construction, then,
the distribution $\rho(q,t)$ of configurations evolves by the continuity
equation%
\begin{equation}
\frac{\partial\rho}{\partial t}+\partial_{q}\cdot\left(  \rho v\right)
=0\ .\label{Cont_rho_gen}%
\end{equation}
This matches the continuity equation (\ref{Cont_psi2_gen}) for $\left\vert
\psi\right\vert ^{2}$. It follows immediately that if $\rho=\left\vert
\psi\right\vert ^{2}$ initially then $\rho=\left\vert \psi\right\vert ^{2}$ at
later times. An ensemble obeying the Born rule is in a state of `quantum
equilibrium'. For such ensembles we recover the usual statistical predictions
of textbook quantum mechanics \cite{B52,Holl93}.

There is, however, no reason of principle why we could not begin with a
`quantum nonequilibrium' ensemble with $\rho\neq\left\vert \psi\right\vert
^{2}$ \cite{AV91a,AV91b,AV92}. What happens then? In general we will find
violations of the usual statistical predictions. For example, for single
particles incident on a two-slit screen with incoming wave function $\psi$, an
incident ensemble with $\rho\neq\left\vert \psi\right\vert ^{2}$ will yield an
anomalous distribution $\rho(\mathbf{x},t)\neq\left\vert \psi(\mathbf{x}%
,t)\right\vert ^{2}$ at the backstop, breaking the usual interference pattern.
Similarly, atomic transitions will have non-standard probabilities, and so on.
And yet, such nonequilibrium phenomena have never been observed in the
laboratory. Why not?

\subsection{Quantum relaxation}

In pilot-wave theory there is a straightforward answer. At some time in the
remote past there took place a process of `quantum relaxation' -- by which we
mean the time evolution of $\rho$ towards $\left\vert \psi\right\vert ^{2}$
(on a coarse-grained level). Quantum equilibrium was already reached, at least
to a very good approximation, long before any of our experiments were carried
out. That is why we see the Born rule today
\cite{AV91a,AV91b,AV92,AV96,AV01,VW05,PV06,AV09,Allori20}. Bearing in mind the
long and violent astrophysical and cosmological history of all known systems,
quantum relaxation probably occurred in the very early universe.

Quantum relaxation can be understood, by analogy with thermal relaxation for
an isolated classical system, in terms of the decrease of a coarse-grained
$H$-function \cite{AV91a,AV92}%
\begin{equation}
\bar{H}(t)=\int dq\ \bar{\rho}\ln(\bar{\rho}/\overline{|\psi|^{2}})\,,
\label{Hqu_cg}%
\end{equation}
where the overbars indicate coarse-graining over small cells in configuration
space. This quantity is equal to minus the relative entropy of $\bar{\rho}$
with respect to $\overline{|\psi|^{2}}$. It is bounded below by zero, $\bar
{H}\geq0$, and $\bar{H}=0$ if and only if $\bar{\rho}=\overline{|\psi|^{2}}$.
If we begin at $t=0$ with $\bar{\rho}\neq\overline{|\psi|^{2}}$ then $\bar
{H}(0)>0$. As the ensemble relaxes towards equilibrium, $\bar{H}%
(t)\rightarrow0$ and $\bar{\rho}\rightarrow\overline{|\psi|^{2}}$. This
relaxing behaviour has been demonstrated in a wide variety of numerical
simulations, yielding an approximately exponential decay
\cite{VW05,TRV12,ACV14}%
\begin{equation}
\bar{H}(t)\approx\bar{H}(0)e^{-t/\tau} \label{exp decay}%
\end{equation}
on a timescale $\tau$ that is (very roughly) comparable to the quantum
timescale $\Delta t=\hslash/\Delta E$ (though $\tau$ also depends on the
coarse-graining length) \cite{TRV12}. Moreover the quantity (\ref{Hqu_cg})
obeys a general coarse-graining $H$-theorem \cite{AV91a,AV92}%
\begin{equation}
\bar{H}(t)\leq\bar{H}(0)\ , \label{subquHthm}%
\end{equation}
assuming no initial fine-grained structure in $\rho$ and $|\psi|^{2}$ at
$t=0$. Closer analysis shows that $\bar{H}(t)$ strictly decreases when $\rho$
develops fine-grained structure -- as tends to happen for velocity fields that
vary over the coarse-graining cells \cite{AV92,Allori20}.

We have said that quantum relaxation probably took place in the early
universe, soon after the big bang. This idea is potentially testable.
According to inflationary cosmology, primordial quantum fluctuations in a
scalar inflaton field were the ultimate source of primordial inhomogeneities,
which later grew by gravitational clumping to form large-scale structure, as
well as seeding the small temperature anisotropies we see today in the cosmic
microwave background (CMB) \cite{LL00,Muk05,PU09}. This means that the
statistical properties of the CMB sky ultimately depend on the Born rule for
quantum field fluctuations in the very early universe. If the Born rule was
broken at sufficiently early times, this could show up as anomalies in the CMB
today \cite{AV07,AV10}. Careful analysis shows that on expanding space quantum
relaxation is suppressed for long-wavelength (super-Hubble)\ field modes,
suggesting that a pre-inflationary era will end with a power deficit at long
wavelengths, which could then carry over to an inflationary phase yielding a
large-scale power deficit in the CMB \cite{AV08a,CV13}. This scenario has been
studied numerically, with some simplifying assumptions \cite{CV15,CV16}. A
large-scale power deficit has in fact been reported in the CMB data
\cite{Planck16}, though its status remains controversial. Fitting the data to
a quantum relaxation model has yielded some tantalising results, but the data
are too noisy for clear conclusions to be drawn \cite{AV10,VPV19}.

\subsection{Trapped forever in quantum death?}

According to pilot-wave theory, at least when correctly interpreted, we are
currently trapped in a state of `quantum death' that is broadly analogous to
the state of classical thermodynamic `heat death' which was much discussed in
the nineteenth century as a seemingly inevitable future end state of our world
(in which all systems have reached the same temperature and it is no longer
possible to convert heat into work) \cite{AV91a,AV91b,AV92,AV96}. According to
pilot-wave theory, a subquantum analogue of the classical heat death
\textit{has already happened} (and a long time ago). In this state, all
systems are subject to the same quantum noise, as described by the Born rule
-- just as, in the classical heat death, all systems are subject to the same
thermal noise (at the same global temperature). In the state of quantum death,
the uncertainty principle prevents us from observing and controlling the
underlying details of de Broglie-Bohm trajectories. As a consequence, we are
unable to control the underlying nonlocal dynamics of entangled systems, and
in particular we are unable to employ entanglement for nonlocal signalling.
But these limitations are not fundamental, they are merely peculiarities of
the state of quantum death (just as the inability to convert heat into work is
a peculiarity of the state of heat death). Locality emerges at the statistical
level only if the Born rule holds exactly. As has been shown explicitly,
nonequilibrium entangled systems generally allow instantaneous signalling
\cite{AV91b,AV02a,AV02b} -- which can be understood as defining a preferred
foliation of spacetime with a global time parameter $t$ \cite{AV08b}.

Here we are interested specifically in the status of the Born rule. Clearly,
in pilot-wave theory, the Born rule is not a law of nature, but holds only
because we are confined to a certain state of statistical equilibrium.
Moreover we seem to be forever trapped in this state, which appears to be
\textit{stable}. The equations of motion of pilot-wave theory guarantee that
once quantum equilibrium is reached there is no way out (barring extremely
rare fluctuations \cite{AV92}, analogous to putting a kettle of water on ice
and waiting for the water to boil). There is one caveat, however. Our
discussion applies to quantum systems on a classical spacetime background.
What happens if spacetime itself is quantised? To answer that, we must turn to
the pilot-wave theory of gravity.

\section{Pilot-wave gravitation}

The pilot-wave theory of gravity appears to have been first written down and
studied for a mini-superspace model by Vink \cite{Vink92} and in general terms
by Horiguchi \cite{Hor94}. It has since been developed, and extensively
applied to cosmology, in particular by Pinto-Neto and collaborators
\cite{PN05,PNF13,PN21}.

Beginning for simplicity with the case of pure gravitation, the Wheeler-DeWitt
equation (\ref{WD1}) for the wave functional $\Psi\lbrack g_{ij}]$ is
supplemented by a de Broglie guidance equation for the time evolution
$g_{ij}(t)$ of the 3-metric,%
\begin{equation}
\frac{\partial g_{ij}}{\partial t}=2NG_{ijkl}\frac{\delta S}{\delta g_{kl}}\ ,
\label{deB QG}%
\end{equation}
where $S$ is the phase of $\Psi$ and for simplicity we have set $N_{i}%
=0$.\footnote{For $N_{i}\neq0$ there are additional terms $D_{i}N_{j}%
+D_{j}N_{i}$ on the right-hand side of (\ref{deB QG}).} The equation of motion
(\ref{deB QG}) can be justified in two ways. We might simply identify the
classical canonical momentum density $p^{ij}$ (conjugate to $g_{ij}$) with the
phase gradient $\delta S/\delta g_{ij}$ and then use the well-known classical
relation between $p^{ij}$ and $\dot{g}_{ij}$ to yield (\ref{deB QG}).
Alternatively, (\ref{deB QG}) can be justified as the natural velocity field
appearing in the equation%
\begin{equation}
\frac{\delta}{\delta g_{ij}}\left(  |\Psi|^{2}G_{ijkl}\frac{\delta S}{\delta
g_{kl}}\right)  =0\ , \label{W-D_cont2}%
\end{equation}
which follows from the Wheeler-DeWitt equation (\ref{WD1}) (with an
appropriate choice of operator ordering,\footnote{Specifically, with the
ordering $(\delta/\delta g_{ij})G_{ijkl}(\delta/\delta g_{kl})$ in the kinetic
term.} inserting $\Psi=\left\vert \Psi\right\vert e^{iS}$ and taking the
imaginary part), and which can be rewritten as%
\begin{equation}
\frac{\delta}{\delta g_{ij}}\left(  |\Psi|^{2}\frac{\partial g_{ij}}{\partial
t}\right)  =0 \label{W-D_cont3a}%
\end{equation}
with $\dot{g}_{ij}$ given by (\ref{deB QG}). Equations (\ref{deB QG}) and
(\ref{WD1}), together with the constraint (\ref{cons1}), are taken to define
the dynamics of a single system.

These equations are readily extended to include a matter field $\phi$. The
Wheeler-DeWitt equation (\ref{W-D_ext}) for $\Psi\lbrack g_{ij},\phi]$ then
includes a matter term (\ref{matter_H}) and $\Psi$ is subject to the
constraint (\ref{cons2}). We still have the same guidance equation
(\ref{deB QG}) for $g_{ij}$ and in addition a guidance equation%
\begin{equation}
\frac{\partial\phi}{\partial t}=\frac{N}{\sqrt{g}}\frac{\delta S}{\delta\phi}
\label{deB_phi}%
\end{equation}
for $\phi$ (again taking $N_{i}=0$).\footnote{For $N_{i}\neq0$ there is an
additional term $N^{i}\partial_{i}\phi$ on the right-hand side of
(\ref{deB_phi}).} As before the guidance equations can be justified by
identifying the classical canonical momenta with a phase gradient, or by
identifying the natural velocity fields appearing in the equation%
\begin{equation}
\frac{\delta}{\delta g_{ij}}\left(  |\Psi|^{2}\frac{\partial g_{ij}}{\partial
t}\right)  +\frac{\delta}{\delta\phi}\left(  |\Psi|^{2}\frac{\partial\phi
}{\partial t}\right)  =0 \label{W-D_cont3b}%
\end{equation}
(which now follows from the extended Wheeler-DeWitt equation (\ref{W-D_ext})).

Before proceeding we should point out that, in the above dynamics, the status
of the spacetime foliation is perhaps not fully understood. The arbitrary
functions $N$, $N^{i}$ should not affect the 4-geometry that is traced out by
the evolving 3-metric (for given initial conditions). Shtanov \cite{Sht96}
argued that the 4-geometry depends on $N$, suggesting that the theory breaks
foliation invariance. In that case one might include a specific choice for $N$
as part of the theory. On the other hand, work by Pinto-Neto and Santini
\cite{PNS02} suggests that the 4-geometry is in fact independent of $N$,
$N^{i}$. By writing the dynamics as a Hamiltonian system, it is argued that
the time evolution of an initial 3-geometry yields the same 4-geometry for all
$N$, $N^{i}$ -- with the caveat that the 4-geometry is non-Lorentzian. Local
Lorentz invariance is broken (as expected in a nonlocal theory). If this
argument is correct it seems to imply that, for a given solution $\Psi$ and
for a given initial 3-geometry, the resulting spacetime has an effective
preferred foliation. Intuitively, this seems consistent with the first-order
(or `Aristotelian') structure of pilot-wave dynamics \cite{AV97}. And, as
already noted, for nonequilibrium ensembles of entangled systems we obtain
statistical nonlocal signals \cite{AV91b,AV02a,AV02b}, which arguably also
define a preferred foliation of spacetime \cite{AV08b}. It would be of
interest to study these matters in more detail.

The above dynamics has been applied extensively by numerous authors, in
particular to quantum cosmology. Such applications have focussed on properties
of the trajectories (such as singularity avoidance) without attempting to
construct a theory of a quantum equilibrium ensemble \cite{PN05,PNF13,PN21}.
In fact previous workers have avoided discussing ensembles, owing to the
pathological (non-normalisable) nature of the density $|\Psi|^{2}$. By a
curious twist, we then find ourselves in a position opposite to that of
textbook quantum mechanics: we have a theory of single systems with
trajectories, but no theory of ensembles or of probabilities. It has been
suggested that this is understandable because (as argued by the Bohmian
mechanics school \cite{DGZ92,DT09}) the notion of probability is (supposedly)
meaningless for a single universe \cite{PN21}. And yet theoretical
cosmologists routinely discuss probabilities for primordial cosmological
perturbations, and observational cosmologists employ measurements of the CMB
to constrain the primordial power spectrum. In practice, by assuming
statistical isotropy and statistical homogeneity for a theoretical ensemble,
we can and do discuss probabilities for our universe and constrain them by
observation \cite{Allori20}. How, then, can we proceed with a theory of
probability in pilot-wave gravitation?

\subsection{Gravity without the Born rule}

Our suggested answer is to accept that at the fundamental level there is no
such thing as the Born rule \cite{AV21}. An arbitrary theoretical ensemble
with the same Wheeler-DeWitt wave functional $\Psi\lbrack g_{ij},\phi]$ will
have an arbitrary initial probability distribution $P[g_{ij},\phi,t_{i}]$ at
time $t_{i}$. By definition $P[g_{ij},\phi,t_{i}]$ will be normalisable and so
cannot be equal to $\left\vert \Psi\lbrack g_{ij},\phi]\right\vert ^{2}$ under
any circumstances. The initial distribution $P[g_{ij},\phi,t_{i}]$ is a
contingency unconstrained by any law but which can, at least in principle, be
constrained by empirical observation. Furthermore, we can straightforwardly
study its time evolution. Each element of the ensemble evolves by the de
Broglie velocity field (\ref{deB QG}) and (\ref{deB_phi}), and so
$P[g_{ij},\phi,t]$ necessarily evolves by the continuity equation\footnote{We
might also append a constraint of the form (\ref{cons2}) on $P$, to ensure
that it is a function on the space of coordinate-independent 3-geometries. We
can avoid this complication by simply working with one representation of the
3-geometry by one (coordinate-dependent) metric $g_{ij}$.}%
\begin{equation}
\frac{\partial P}{\partial t}+\int d^{3}x\ \frac{\delta}{\delta g_{ij}}\left(
P\frac{\partial g_{ij}}{\partial t}\right)  +\int d^{3}x\ \frac{\delta}%
{\delta\phi}\left(  P\frac{\partial\phi}{\partial t}\right)  =0\ .
\label{Cont P QG}%
\end{equation}

We then have a theory for a general ensemble of gravitational systems evolving
in time. One of our key claims is that this theory has no Born-rule
equilibrium state \cite{AV21,AV14}. At the deepest level of gravitational
physics, the universe is in a perpetual state of quantum nonequilibrium%
\begin{equation}
P[g_{ij},\phi,t]\neq\left\vert \Psi\lbrack g_{ij},\phi]\right\vert ^{2}\ .
\end{equation}
In Section 6 we shall illustrate these ideas with our simple model of quantum cosmology.

\subsection{Naive Schr\"{o}dinger interpretation. II}

Some workers instead interpret $\left\vert \Psi\right\vert ^{2}$ as a
probability density (as first suggested by Vink \cite{Vink92} and recently
advocated by D\"{u}rr and Struyve \cite{DS20}). This amounts to applying the
naive Schr\"{o}dinger interpretation to pilot-wave gravitation. However, while
the presence of trajectories adds a new element, the interpretation remains
unworkable because $\left\vert \Psi\right\vert ^{2}$ is non-normalisable.

It might be thought that equation (\ref{W-D_cont3b}) can be employed to
motivate $\left\vert \Psi\right\vert ^{2}$ is an equilibrium probability
density. But (\ref{W-D_cont3b}) is not a continuity equation but an infinity
of equations (one per spatial point $x$). Following D\"{u}rr and Struyve
\cite{DS20}, if we integrate (\ref{W-D_cont3b}) over $x$ we can write down
what we call a `pseudo-continuity equation'\footnote{For a single particle
with a static density $\rho$ and a current $\mathbf{j}$, this is analogous to
summing the equations $\partial_{x}j_{x}=\partial_{y}j_{y}=\partial_{z}%
j_{z}=0$ to yield $\partial\rho/\partial t+\mathbf{\nabla}\cdot\mathbf{j}=0$.}%
\begin{equation}
\frac{\partial\left\vert \Psi\right\vert ^{2}}{\partial t}+\int d^{3}%
x\ \frac{\delta}{\delta g_{ij}}\left(  \left\vert \Psi\right\vert ^{2}%
\frac{\partial g_{ij}}{\partial t}\right)  +\int d^{3}x\ \frac{\delta}%
{\delta\phi}\left(  \left\vert \Psi\right\vert ^{2}\frac{\partial\phi
}{\partial t}\right)  =0 \label{pseudo_cont1}%
\end{equation}
(where for completeness we have inserted the vanishing term $\partial
\left\vert \Psi\right\vert ^{2}/\partial t=0$). This is formally the same as
the physical continuity equation (\ref{Cont P QG}) for $P$. We might then
`deduce' that $P=\left\vert \Psi\right\vert ^{2}$ is an equilibrium state, as
usually done for non-gravitational systems. On this basis D\"{u}rr and Struyve
claim that $\left\vert \Psi\right\vert ^{2}$ can be employed as a quantum
equilibrium measure of `typicality' for initial configurations of the universe
(from which follows the Born rule for subsystems, along lines already
advocated by the Bohmian mechanics school). But this novel application of the
naive Schr\"{o}dinger interpretation again founders on the fact that
$\left\vert \Psi\right\vert ^{2}$ is not normalisable and cannot define a
physical probability (or typicality) measure.

We should be wary of artificial attempts to make $\left\vert \Psi\right\vert
^{2}$ appear like a conventional density. In our view, to interpret
$\left\vert \Psi\right\vert ^{2}$ as a Born-rule measure is a category
mistake. At the fundamental level there is no Born rule. As we will see the
usual Born-rule measure emerges only in the Schr\"{o}dinger approximation (on
a classical spacetime background).

\section{Pilot-wave cosmology}

Recall our quantum-cosmological model with degrees of freedom $(a,\phi)$ and
wave function $\Psi(a,\phi)$ satisfying the Wheeler-DeWitt equation
(\ref{WD_mini}). Inserting $\Psi=\left\vert \Psi\right\vert e^{iS}$ and taking
the imaginary part yields what we call a `pseudo-continuity equation'%
\begin{equation}
\frac{\partial}{\partial a}\left(  a^{2}\left\vert \Psi\right\vert ^{2}\dot
{a}\right)  +\frac{\partial}{\partial\phi}\left(  a^{2}\left\vert
\Psi\right\vert ^{2}\dot{\phi}\right)  =0 \label{p-cont_qucosmo_t}%
\end{equation}
for a density $a^{2}\left\vert \Psi\right\vert ^{2}$ and with a velocity field%
\begin{equation}
\dot{a}=-\frac{1}{m_{\mathrm{P}}^{2}}\frac{1}{a}\frac{\partial S}{\partial
a}\ ,\ \ \dot{\phi}=\frac{1}{a^{3}}\frac{\partial S}{\partial\phi}\ .
\label{deB_qu_cosm_t}%
\end{equation}
We can identify (\ref{deB_qu_cosm_t}) as the natural de Broglie guidance
equations for this system.\footnote{The same equations follow by identifying
the canonical momenta (\ref{can_moma_qucosmo}) with a phase gradient.}

A general theoretical ensemble of systems with the same wave function $\Psi$
will have a probability distribution $P(a,\phi,t)$ (with density defined with
respect to $dad\phi$). Since each element of the ensemble evolves according to
the velocity field (\ref{deB_qu_cosm_t}), the distribution $P(a,\phi,t)$
necessarily evolves according to the continuity equation%
\begin{equation}
\frac{\partial P}{\partial t}+\frac{\partial}{\partial a}\left(  P\dot
{a}\right)  +\frac{\partial}{\partial\phi}\left(  P\dot{\phi}\right)  =0
\label{ContP_qucosmo_t}%
\end{equation}
(with $\dot{a}$, $\dot{\phi}$ given by (\ref{deB_qu_cosm_t})). We then have a
theory for a general ensemble of cosmological systems evolving in time.

\subsection{Failure of the Born rule}

At this point we must be careful. In standard quantum theory equation
(\ref{p-cont_qucosmo_t}) would be regarded as a continuity equation for a
Born-rule-like probability density $a^{2}\left\vert \Psi\right\vert ^{2}$. The
unusual factor $a^{2}$ arises simply from the structure of our minisuperspace
and if preferred could be eliminated by defining a density with respect to
$a^{2}dad\phi$ instead of $dad\phi$. In any case, equation
(\ref{p-cont_qucosmo_t}) seemingly suggests that the quantum probability to
find the system in a range $dad\phi$ is given by $a^{2}\left\vert
\Psi\right\vert ^{2}dad\phi$. This is the naive Schr\"{o}dinger interpretation
in a quantum-cosmological setting. As we saw in Section 3.2 this
interpretation fails not only because the putative probability density
$a^{2}\left\vert \Psi\right\vert ^{2}$ has no explicit time dependence but
also because it is non-normalisable.

In pilot-wave theory it might be argued that, because the respective evolution
equations (\ref{ContP_qucosmo_t}) and (\ref{p-cont_qucosmo_t}) for $P$ and
$a^{2}\left\vert \Psi\right\vert ^{2}$ are identical (noting that
$\partial(a^{2}\left\vert \Psi\right\vert ^{2})/\partial t=0$), if
$P=a^{2}\left\vert \Psi\right\vert ^{2}$ initially then $P=a^{2}\left\vert
\Psi\right\vert ^{2}$ at later times and we may identify this as a state of
quantum equilibrium. This is again the naive Schr\"{o}dinger interpretation
applied to pilot-wave theory (Section 5.2), and again it is as untenable in
pilot-wave theory as it is in standard quantum theory. For a general solution
$\Psi$ of the wave equation (\ref{WD_mini}) we inevitably have%
\begin{equation}
\int\int dad\phi\ a^{2}\left\vert \Psi(a,\phi)\right\vert ^{2}=\infty\ .
\end{equation}
The equality $P=a^{2}\left\vert \Psi\right\vert ^{2}$ is mathematically
nonsensical since the left-hand side is (by definition) normalisable while the
right-hand side is not.

To understand this theory we need to accept that in the deep quantum-gravity
regime there is \textit{no} Born-rule-like equilibrium state. There simply is
no Born rule and no state of quantum equilibrium. In the context of our
quantum-cosmological model, we must have%
\begin{equation}
P\neq a^{2}\left\vert \Psi\right\vert ^{2}%
\end{equation}
always (both initially and at later times).

\subsection{Impossibility of quantum relaxation}

In terms of quantum relaxation, for this system the coarse-grained
$H$-function%
\begin{equation}
\bar{H}(t)=\int\int dad\phi\ \bar{P}\ln(\bar{P}/\overline{a^{2}\left\vert
\Psi\right\vert ^{2}}) \label{Hbar_QC}%
\end{equation}
(minus the relative entropy of $\bar{P}$ with respect to $\overline
{a^{2}\left\vert \Psi\right\vert ^{2}}$) still obeys a coarse-graining
$H$-theorem (\ref{subquHthm}) but now has no lower bound. If we had%
\begin{equation}
\int\int dad\phi\ \overline{a^{2}\left\vert \Psi\right\vert ^{2}}=N
\end{equation}
for some finite $N$, it is easy to show that (\ref{Hbar_QC}) would be bounded
below by $-\ln N$, with the lower bound attained if and only if $\bar{P}%
=\frac{1}{N}\overline{a^{2}\left\vert \Psi\right\vert ^{2}}$ \cite{AV21}. For
$N\rightarrow\infty$ there is no lower bound. The function $\bar{H}(t)$ can
decrease indefinitely without ever reaching a minimum. In this sense $\bar{P}$
is always infinitely far away from the putative `equilibrium' state
$\overline{a^{2}\left\vert \Psi\right\vert ^{2}}$. Limited local relaxation
might take place in some regions of configuration space, but to attain global
equilibrium is mathematically impossible. Similar reasoning applies to the
full gravitational theory \cite{AV21}.

\section{Quantum relaxation in the Schr\"{o}dinger approximation}

We have argued that, in the deep quantum-gravity regime, quantum relaxation
cannot take place because there is no physical equilibrium state to relax to.
In effect a quantum-gravitational system is perpetually in nonequilibrium. How
are we then to understand the ubiquity of the Born rule in our world today?

Our proposed answer is that quantum relaxation can take place in the
semiclassical regime, where the system propagates on an approximately
classical spacetime background. In this `Schr\"{o}dinger approximation' we
have an effective time-dependent Schr\"{o}dinger equation (\ref{Sch_gen}) for
a conventional wave function $\psi(q,t)$, where $q$ might represent for
example a field configuration $\phi$ on a background classical curved space
and $t$ is the time function associated with a preferred foliation. Since
$\psi$ is now normalisable, $\left\vert \psi\right\vert ^{2}$ can correspond
to a physical probability distribution, which is attainable after appropriate relaxation.

To see how this works, we need to outline how the Schr\"{o}dinger
approximation is derived from the underlying quantum-gravitational theory
(details are given in Section 9.1). Consider again the deep quantum-gravity
regime with a matter field $\phi$ and a Wheeler-DeWitt wave functional
$\Psi\lbrack g_{ij},\phi]$. We obtain an effective time-dependent wave
function $\psi$, on an approximately classical spacetime background, when the
solution $\Psi\lbrack g_{ij},\phi]$ of the Wheeler-DeWitt equation takes the
approximate form%
\begin{equation}
\Psi\lbrack g_{ij},\phi]\approx\Psi_{\mathrm{WKB}}[g_{ij}]\psi\lbrack
\phi,g_{ij}]\ , \label{WKB}%
\end{equation}
where $\Psi_{\mathrm{WKB}}[g_{ij}]$ is a WKB wave functional for the 3-metric.
The phase $S_{\mathrm{WKB}}=\operatorname{Im}\ln\Psi_{\mathrm{WKB}}$ satisfies
a classical Hamilton-Jacobi equation and generates classical trajectories
$g_{ij}=g_{ij}(t)$ for the background. If we evaluate $\psi\lbrack\phi
,g_{ij}]$ along a specific trajectory $g_{ij}(t)$, we can define an effective
time-dependent wave functional%
\begin{equation}
\psi_{\mathrm{eff}}[\phi,t]=\psi\lbrack\phi,g_{ij}(t)]
\end{equation}
for the matter field $\phi$, with a time derivative%
\begin{equation}
\frac{\partial}{\partial t}=\int d^{3}x\ \dot{g}_{ij}\frac{\delta}{\delta
g_{ij}}\ . \label{WKBt}%
\end{equation}
It can then be shown that $\psi_{\mathrm{eff}}$ satisfies an approximate
time-dependent Schr\"{o}dinger equation%
\begin{equation}
i\frac{\partial\psi_{\mathrm{eff}}}{\partial t}=\hat{H}_{\mathrm{eff}}%
\psi_{\mathrm{eff}} \label{Sch_eff}%
\end{equation}
for the field $\phi$ on the classical background, where $\hat{H}%
_{\mathrm{eff}}$ is an effective Hamiltonian (which of course depends on the background).

This method of deriving the Schr\"{o}dinger approximation has a long history.
The WKB trajectories for the classical background allow us to define an
effective time parameter $t$, which historically has often been called `WKB
time' \cite{Zeh88}. The origin of such trajectories is unclear in standard
quantum mechanics, where they are really being inserted by hand. In pilot-wave
theory, in contrast, the WKB trajectories are simply de Broglie-Bohm
trajectories evaluated in the WKB approximation, and so the above construction
is conceptually clear.

We can now return to the question of quantum relaxation and the Born rule.
Once the very early universe enters the semiclassical or Schr\"{o}dinger
regime, fields and particles propagating on the (approximate) classical
background will satisfy a time-dependent Schr\"{o}dinger equation of the form
(\ref{Sch_eff}), with a conventional and normalisable wave function
$\psi_{\mathrm{eff}}$. As we will see in Section 9.1, the de Broglie guidance
equation also takes the standard form (in terms of $\psi_{\mathrm{eff}}$). We
then find ourselves in the domain which has already been much studied in
pilot-wave theory as briefly summarised in Section 4.1. If at the beginning of
the Schr\"{o}dinger regime we have a nonequilibrium probability distribution
$\rho\neq\left\vert \psi_{\mathrm{eff}}\right\vert ^{2}$, then in appropriate
circumstances quantum relaxation will ensure that $\rho\rightarrow\left\vert
\psi_{\mathrm{eff}}\right\vert ^{2}$ on a coarse-grained level, at least to a
good approximation and in particular for short-wavelength (sub-Hubble) field
modes. In this way, despite the complete absence of a Born rule in the deep
quantum-gravity regime, we can nevertheless understand the emergence of the
Born rule in the semiclassical or Schr\"{o}dinger approximation, at scales
relevant to laboratory physics, after appropriate quantum relaxation.

We have said that, once we have an approximate time-dependent Schr\"{o}dinger
equation, conventional quantum relaxation can take place. But is there any
reason to expect nonequilibrium $\rho\neq\left\vert \psi_{\mathrm{eff}%
}\right\vert ^{2}$ at the start of the semiclassical regime? Indeed there is.
Fundamentally we have a perpetual nonequilibrium ensemble with distribution
$P[g_{ij},\phi,t]\neq\left\vert \Psi\lbrack g_{ij},\phi]\right\vert ^{2}$. As
we enter the semiclassical regime, say at some `initial' time $t_{i}$
(approximately marking the beginning of that regime), the field $\phi$ will
have a conditional probability density%
\begin{equation}
\rho\lbrack\phi,t_{i}]=\frac{P[g_{ij},\phi,t_{i}]}{\left(  \int P[g_{ij}%
,\phi,t_{i}]D\phi\right)  }\ , \label{condnl_prob_phi}%
\end{equation}
where on the right-hand side it is understood that we have inserted the actual
value of the classical background 3-metric $g_{ij}$ at time $t_{i}$. Because
here%
\begin{equation}
P[g_{ij},\phi,t_{i}]\neq\left\vert \Psi\lbrack g_{ij},\phi]\right\vert
^{2}\approx\left\vert \Psi_{\mathrm{WKB}}[g_{ij}]\right\vert ^{2}\left\vert
\psi\lbrack\phi,g_{ij}]\right\vert ^{2}=\left\vert \Psi_{\mathrm{WKB}}%
[g_{ij}]\right\vert ^{2}\left\vert \psi_{\mathrm{eff}}[\phi,t_{i}]\right\vert
^{2}\ ,
\end{equation}
it follows that%
\begin{equation}
\rho\lbrack\phi,t_{i}]\neq\left\vert \psi_{\mathrm{eff}}[\phi,t_{i}%
]\right\vert ^{2} \label{noneq_Sch}%
\end{equation}
(unless it so happens that $P[g_{ij},\phi,t_{i}]=\Pi\lbrack g_{ij}]\left\vert
\psi_{\mathrm{eff}}[\phi,t_{i}]\right\vert ^{2}$ for some $\Pi\lbrack g_{ij}%
]$). We then expect to find quantum nonequilibrium at the start of the
semiclassical or Schr\"{o}dinger regime, with the Born rule emerging only
later after an appropriate period of quantum relaxation.

\section{Instability of the Born rule in quantum gravity}

We have outlined how the time-dependent Schr\"{o}dinger equation
(\ref{Sch_eff}) for the effective wave function $\psi_{\mathrm{eff}}$ emerges
in the semiclassical approximation. We then have a normalisable wave function
and quantum relaxation to equilibrium $\rho\rightarrow\left\vert
\psi_{\mathrm{eff}}\right\vert ^{2}$ can proceed in the usual way. The
Schr\"{o}dinger equation (\ref{Sch_eff}) is, however, subject to small
quantum-gravitational corrections to the effective Hamiltonian $\hat
{H}_{\mathrm{eff}}$. Remarkably, some of the correction terms are
non-Hermitian \cite{KS91,KK12,Bini13,BKM16}. These terms have no consistent
interpretation in standard quantum mechanics as they violate the conservation
of probability. In pilot-wave theory, in contrast, probability is by
construction conserved and (as we shall see) the non-Hermitian terms simply
render the Born rule unstable.

The derivation of the correction terms will be presented in the next section.
Here we first show how pilot-wave theory is able to accomodate such terms consistently.

The corrections are calculated by performing a `semiclassical expansion' of
the Wheeler-DeWitt equation (Section 9). Dropping for simplicity the subscript
`\textrm{eff}', we find an effective Hamiltonian of the form%
\begin{equation}
\hat{H}=\hat{H}_{\phi}+\hat{H}_{a}+i\hat{H}_{b}\ , \label{H_eff}%
\end{equation}
where $\hat{H}_{\phi}$ is the usual field Hamiltonian and the Hermitian
operators $\hat{H}_{a}$, $\hat{H}_{b}$ represent tiny quantum-gravitational
corrections. There is a Hermitian correction $\hat{H}_{a}$ and a non-Hermitian
correction $i\hat{H}_{b}$. Writing $\hat{H}_{1}=\hat{H}_{\phi}+\hat{H}_{a}$
and $\hat{H}_{2}=\hat{H}_{b}$, the effective Schr\"{o}dinger equation for
$\psi\lbrack\phi,t]$ takes the form%
\begin{equation}
i\frac{\partial\psi}{\partial t}=(\hat{H}_{1}+i\hat{H}_{2})\psi\ .
\label{Sch_nonH}%
\end{equation}
Applying the same semiclassical expansion to the de Broglie guidance equation,
we find an effective guidance equation of the form%
\begin{equation}
\frac{\partial\phi}{\partial t}=\frac{j_{1}}{\left\vert \psi\right\vert ^{2}%
}\ , \label{deB_nonH}%
\end{equation}
where $j_{1}$ is the usual current associated with the Hermitian part $\hat
{H}_{1}$ only. Thus, while the Schr\"{o}dinger equation (\ref{Sch_nonH}) has a
non-Hermitian correction $i\hat{H}_{2}$, this does not affect the guidance
equation (\ref{deB_nonH}). As we show in the next section, these results
follow directly and without ambiguity from the underlying
quantum-gravitational equations in a semiclassical expansion.

To see the consequences note that (\ref{Sch_nonH}) implies a continuity
equation (writing $j_{1}=\left\vert \psi\right\vert ^{2}\dot{\phi}$)%
\begin{equation}
\frac{\partial\left\vert \psi\right\vert ^{2}}{\partial t}+\partial_{\phi
}\cdot(\left\vert \psi\right\vert ^{2}\dot{\phi})=s\ , \label{Cont_psi2_s}%
\end{equation}
where $\partial_{\phi}\cdot(...)=\int d^{3}x\ \delta/\delta\phi(x)\ (...)$ is
a divergence in field configuration space and%
\begin{equation}
s=2\operatorname{Re}\left(  \psi^{\ast}\hat{H}_{2}\psi\right)  \ . \label{s}%
\end{equation}
For an ensemble of systems with the same wave function $\psi$, each element of
the ensemble evolves by the de Broglie velocity field (\ref{deB_nonH}). The
probability density $\rho\lbrack\phi,t]$ then evolves by the usual continuity
equation%
\begin{equation}
\frac{\partial\rho}{\partial t}+\partial_{\phi}\cdot(\rho\dot{\phi})=0\ .
\label{Cont_rho}%
\end{equation}
For $s\neq0$ there is a mismatch between equations (\ref{Cont_psi2_s}) and
(\ref{Cont_rho}). It follows that an initial distribution $\rho=|\psi|^{2}$
can evolve into a final distribution $\rho\neq|\psi|^{2}$. The Born rule is
unstable. This can be quantified in terms of the ratio $f=\rho/|\psi|^{2}$,
which is no longer conserved along trajectories. From (\ref{Cont_psi2_s}) and
(\ref{Cont_rho}) we find%
\begin{equation}
\frac{df}{dt}=-\frac{sf}{|\psi|^{2}}\ \label{dfdt}%
\end{equation}
(where $d/dt=\partial/\partial t+\int d^{3}x\ \dot{\phi}.\delta/\delta\phi(x)$
is the time derivative along a trajectory in field configuration space). It is
also worth noting that the squared-norm $\int D\phi\ \left\vert \psi
\right\vert ^{2}$ of $\psi$ changes with time,%
\begin{equation}
\frac{d}{dt}\int D\phi\ \left\vert \psi\right\vert ^{2}=\int
dq\ s=2\left\langle \hat{H}_{2}\right\rangle \ .
\end{equation}

We see from the above equations that the usual Born-rule equilibrium state
$\rho=|\psi|^{2}$ is unstable. As a result of quantum-gravitational
corrections, nonequilibrium $\rho\neq|\psi|^{2}$ is created on a timescale
$\tau_{\mathrm{noneq}}$ which can be estimated from the rate of change of the
(fine-grained) $H$-function $H(t)=\int D\phi\ \rho\ln(\rho/\left\vert
\psi\right\vert ^{2})$. From (\ref{Cont_psi2_s}) and (\ref{Cont_rho}) we
find\footnote{When $s=0$ the exact $H$ is constant but the coarse-grained
value decreases (if there is no initial fine-grained structure) \cite{AV91a}.}%
\begin{equation}
\frac{dH}{dt}=-\int D\phi\ \frac{\rho}{\left\vert \psi\right\vert ^{2}}s\ .
\end{equation}
Close to equilibrium ($\rho\approx\left\vert \psi\right\vert ^{2}$) we have%
\begin{equation}
\frac{dH}{dt}\approx-\int D\phi\ s=-2\left\langle \hat{H}_{2}\right\rangle \ .
\end{equation}
Defining $\tau_{\mathrm{noneq}}$ as the timescale over which $H$ changes by a
factor of order unity, we have the estimate%
\begin{equation}
\tau_{\mathrm{noneq}}\approx\frac{1}{2\left\vert \left\langle \hat{H}%
_{2}\right\rangle \right\vert }\ . \label{tau_est}%
\end{equation}

Note however that, for such effects to build up over time, nonequilibrium must
be created faster than relaxation can remove it, which requires conditions
where%
\begin{equation}
\tau_{\mathrm{relax}}>\tau_{\mathrm{noneq}}\ . \label{condn}%
\end{equation}

Some quantum-gravity theorists have long been puzzled by the non-Hermitian
terms $i\hat{H}_{2}$ (first found in 1991 and re-derived in more recent
papers), which signal a violation of unitarity (the usual norm of $\psi$ is
not conserved) \cite{KS91,KK12,Bini13,BKM16}. Because the non-Hermitian terms
are inconsistent with the standard interpretation of $\left\vert
\psi\right\vert ^{2}$ as a probability density, they are often regarded as an
artifact to be ignored by fiat. Some authors have advocated formally
eliminating these terms by appropriate redefinitions of the wave function
\cite{Bini13,KW18}. We suggest that such redefinitions may turn out to be an
artificial means of disguising genuine physical effects. Our experience with
quantum systems on a classical spacetime background teaches us that the
Hamiltonian must be Hermitian, but that experience is limited to conditions
where quantum-gravitational effects are negligible. As we have seen,
non-Hermitian terms are perfectly consistent with pilot-wave theory, according
to which they simply generate a gravitational instability of the Born rule: an
initial density $\rho=\left\vert \psi\right\vert ^{2}$ can evolve into a final
density $\rho\neq\left\vert \psi\right\vert ^{2}$.

The derivation of the non-Hermitian terms will now be discussed in more detail.

\section{Semiclassical expansion}

Quantum-gravitational corrections to the Schr\"{o}dinger equation
(\ref{Sch_eff}) were derived by Kiefer and Singh \cite{KS91} from a
semiclassical expansion%
\begin{equation}
\Psi=\exp i\left(  \mu S_{0}+S_{1}+\mu^{-1}S_{2}+...\right)  \label{semicl}%
\end{equation}
of the extended Wheeler-DeWitt equation (\ref{W-D_ext}) for $\Psi\lbrack
g_{ij},\phi]$, where $\mu=c^{2}/32\pi G$ (dimensions mass per length).
Inserting (\ref{semicl}) into the left-hand side of (\ref{W-D_ext}), terms of
the same order in $\mu$ are collected and their sum set to zero. The orders
that appear are $\mu^{2}$, $\mu$, $\mu^{0}$, $\mu^{-1}$ ... .

To a first approximation we obtain the usual Schr\"{o}dinger equation
(\ref{Sch_eff}) for a field $\phi$ on a classical background spacetime. We
then obtain gravitational corrections to (\ref{Sch_eff}), in the form of (very
small) Hermitian and non-Hermitian terms in the Hamiltonian. The results found
by Kiefer and Singh are summarised below. In pilot-wave theory we must also
consider how the semiclassical expansion affects the de Broglie guidance
equation (\ref{deB_phi}) for $\phi$. As we will see, the guidance equation
retains its standard form. Thus, if the semiclassical expansion is to be
trusted, it follows from the fundamental equations of quantum gravity that the
emergent Born rule is unstable.

\subsection{Lowest-order Schr\"{o}dinger approximation}

At order $\mu^{2}$ the expansion (\ref{semicl}) yields $\left(  \delta
S_{0}/\delta\phi\right)  ^{2}=0$ so that $S_{0}=S_{0}[g_{ij}]$ depends only on
$g_{ij}$, while at order $\mu$ it is found that $S_{0}$ satisfies a classical
Hamilton-Jacobi equation whose solution defines a classical background
spacetime. The trajectories of the classical background can be used to define
an effective time parameter $t$ (cf. equation (\ref{WKBt})).

Order $\mu^{0}$ yields an equation for $S_{1}$, which can be written as an
effective time-dependent Schr\"{o}dinger equation for a zeroth-order
(uncorrected) wave functional $\psi^{(0)}[\phi,t]$ on a classical background
with metric $g_{ij}$. Defining%
\begin{equation}
\psi^{(0)}=D\exp(iS_{1}) \label{psi0}%
\end{equation}
for an appropriate functional $D[g_{ij}]$, it can be shown that%
\begin{equation}
i\frac{\partial\psi^{(0)}}{\partial t}=\int d^{3}x\ N\mathcal{\hat{H}}_{\phi
}\psi^{(0)} \label{Sch_0}%
\end{equation}
where $\mathcal{\hat{H}}_{\phi}$ is given by (\ref{matter_H}). This is the
standard Schr\"{o}dinger equation for a massless (minimally-coupled) real
scalar field $\phi$ with potential $\mathcal{V}(\phi)$ on a classical
spacetime background.

As expected, to this order the Wheeler-DeWitt wave functional $\Psi\lbrack
g_{ij},\phi]$ takes the WKB form (\ref{WKB}), with $\Psi_{\mathrm{WKB}}%
[g_{ij}]=(1/D)\exp\left(  iMS_{0}\right)  $ and $\psi\lbrack\phi,g_{ij}%
]=\psi^{(0)}[\phi,t]$.

In pilot-wave theory we must also consider the de Broglie guidance equation
(\ref{deB_phi}) for $\phi$. To this order, how is the field velocity
$\dot{\phi}$ related to the effective wave functional $\psi^{(0)}[\phi,t]$? We
can find out by inserting the expansion (\ref{semicl}) into (\ref{deB_phi})
(where $S=\operatorname{Im}\ln\Psi$), yielding%
\begin{equation}
\frac{\partial\phi}{\partial t}=\frac{N}{\sqrt{g}}\frac{\delta}{\delta\phi
}\left(  \operatorname{Re}S_{1}+\mu^{-1}\operatorname{Re}S_{2}+...\right)  \ .
\label{phidot2}%
\end{equation}
The factor $D$ in (\ref{psi0}) can be chosen to be real, so that
$\operatorname{Re}S_{1}$ is equal to the phase of $\psi^{(0)}$. To lowest
order we then have a de Broglie velocity,%
\begin{equation}
\left(  \frac{\partial\phi}{\partial t}\right)  ^{(0)}=\frac{N}{\sqrt{g}}%
\frac{\delta S^{(0)}}{\delta\phi}\ , \label{deB_0}%
\end{equation}
where $S^{(0)}=\operatorname{Im}\ln\psi^{(0)}$ is the phase of $\psi^{(0)}%
$.\footnote{This result is of course expected from the WKB form (\ref{WKB})
(with $\psi=\psi^{(0)}$).} This is the standard de Broglie guidance equation
for a field $\phi$ with wave functional $\psi^{(0)}[\phi,t]$ \cite{AV04a}.

Thus, in this approximation we recover the usual pilot-wave dynamics of a
field on a classical spacetime background, and so we can expect quantum
relaxation to the Born rule to occur in the usual way.

\subsection{Gravitational corrections}

Following ref. \cite{KS91} we now consider higher orders in the semiclassical
expansion (\ref{semicl}). At order $\mu^{-1}$ Kiefer and Singh obtain an
equation for $S_{2}$. Writing $S_{2}=\sigma_{2}[g_{ij}]+\eta\lbrack\phi
,g_{ij}]$ for appropriately chosen $\sigma_{2}$, the corrected matter wave
functional%
\begin{equation}
\psi^{(1)}=\psi^{(0)}\exp(i\eta/\mu) \label{psi_(1)}%
\end{equation}
is found to satisfy a corrected Schr\"{o}dinger equation%
\begin{equation}
i\frac{\partial\psi^{(1)}}{\partial t}=\int d^{3}x\ N\left(  \mathcal{\hat{H}%
}_{\phi}+\mathcal{\hat{H}}_{a}+i\mathcal{\hat{H}}_{b}\right)  \psi^{(1)}\ ,
\label{Sch_correcns}%
\end{equation}
where%
\begin{equation}
\mathcal{\hat{H}}_{a}=\frac{1}{8\mu}\frac{1}{\sqrt{g}R}\mathcal{\hat{H}}%
_{\phi}^{2} \label{Sch_correcns_a}%
\end{equation}
and%
\begin{equation}
\mathcal{\hat{H}}_{b}=\frac{1}{8\mu}\frac{\delta}{\delta\tau}\left(
\frac{\mathcal{\hat{H}}_{\phi}}{\sqrt{g}R}\right)  \label{Sch_correcns_b}%
\end{equation}
are both Hermitian (employing the convenient shorthand $\delta/\delta\tau
=\dot{g}_{ij}\delta/\delta g_{ij}$, with $\dot{g}_{ij}=2NG_{ijkl}\delta
S_{0}/\delta g_{kl}$, to denote a `many-fingered time derivative' on the
background). To this order we have a total effective Hamiltonian of the form
(\ref{H_eff}) with%
\begin{equation}
\hat{H}_{\phi}=\int d^{3}x\ N\mathcal{\hat{H}}_{\phi}\ ,\ \ \ \hat{H}_{a}=\int
d^{3}x\ N\mathcal{\hat{H}}_{a}\ ,\ \ \ \hat{H}_{b}=\int d^{3}x\ N\mathcal{\hat
{H}}_{b}\ .
\end{equation}
As noted we have Hermitian and non-Hermitian corrections $\hat{H}_{a}$ and
$i\hat{H}_{b}$ respectively.\footnote{In an expanding cosmological background
the ratio of $i\hat{H}_{b}$ to $\hat{H}_{a}$ is roughly of order $\sim H/E$,
where $H=\dot{a}/a$ is the Hubble parameter and $E$ is a typical energy for
the field \cite{KS91}.}

We can now consider the next order in the semiclassical expansion
(\ref{semicl}) of the de Broglie guidance equation (\ref{deB_phi}). Because
the term $\sigma_{2}[g_{ij}]$ in $S_{2}$ is independent of $\phi$, the de
Broglie velocity (\ref{phidot2}) takes the form%
\begin{equation}
\frac{\partial\phi}{\partial t}=\frac{N}{\sqrt{g}}\frac{\delta}{\delta\phi
}\left(  \operatorname{Re}S_{1}+\mu^{-1}\operatorname{Re}\eta+...\right)  \ .
\label{phidot3}%
\end{equation}
The corrected wave functional (\ref{psi_(1)}) has a total phase%
\begin{equation}
\operatorname{Im}\ln\psi^{(1)}=\operatorname{Re}S_{1}+\mu^{-1}%
\operatorname{Re}\eta\ ,
\end{equation}
and so the corrected de Broglie velocity (\ref{phidot3}) can once again be
written in the standard form,%
\begin{equation}
\left(  \frac{\partial\phi}{\partial t}\right)  ^{(1)}=\frac{N}{\sqrt{g}}%
\frac{\delta S^{(1)}}{\delta\phi}\ , \label{deB_no correcn}%
\end{equation}
where now $S^{(1)}=\operatorname{Im}\ln\psi^{(1)}$ is the phase of $\psi
^{(1)}$.

To conclude, despite the non-Hermitian term in the corrected Schr\"{o}dinger
equation (\ref{Sch_correcns}), the de Broglie velocity (\ref{deB_no correcn})
continues to take the standard form (now in terms of $\psi^{(1)}$). In other
words, the expression for the velocity remains that associated with the
original (uncorrected) Hermitian part $\hat{H}_{\phi}$ of the Hamiltonian. The
non-Hermitian term affects the time evolution of $\psi^{(1)}$ -- and so
indirectly affects the trajectories -- but does not change the form of the
guidance equation itself. As we have seen, this implies that the Born rule for
$\phi$ is unstable.

More recently, Brizuela, Kiefer and Kr\"{a}mer \cite{BKM16} have derived
similar results for a minisuperspace model of quantum cosmology. The classical
background is defined by a scale factor $a(t)$ and a homogeneous field
$\phi(t)$. Quantum scalar perturbations (of the background metric combined
with the inflaton perturbation) are described by the Mukhanov-Sasaki variable
$\upsilon_{\mathbf{k}}$ in Fourier space. The wave function $\Psi_{\mathbf{k}%
}(a,\phi,\upsilon_{\mathbf{k}})$ satisfies a Wheeler-DeWitt equation for the
mode $\mathbf{k}$, which is solved by means of a semiclassical expansion%
\begin{equation}
\Psi_{\mathbf{k}}=\exp\left[  i\left(  m_{\mathrm{P}}^{2}S_{0}+m_{\mathrm{P}%
}^{0}S_{1}+m_{\mathrm{P}}^{-2}S_{2}+...\right)  \right]  \label{Psi_k_WKB}%
\end{equation}
in powers of $m_{\mathrm{P}}^{2}$. Inserting this into the left-hand side of
the Wheeler-DeWitt equation, terms of the same order in $m_{\mathrm{P}}$ are
collected and their sum set to zero. By this means, Brizuela et al. derive a
Schr\"{o}dinger equation for an effective wave function $\psi_{\mathbf{k}%
}^{(1)}=\psi_{\mathbf{k}}^{(1)}(\upsilon_{\mathbf{k}},t)$, where the
corrections in the effective Hamiltonian have both Hermitian and non-Hermitian
parts. The same expansion can again be applied to the de Broglie guidance
equation for the perturbations $\upsilon_{\mathbf{k}}$ \cite{AV21}. We again
find that the de Broglie velocity takes the standard form proportional to the
gradient of the phase $s_{\mathbf{k}}^{(1)}=\operatorname{Im}\ln
\psi_{\mathbf{k}}^{(1)}$ of $\psi_{\mathbf{k}}^{(1)}$, and so remains equal to
the velocity generated by the (uncorrected) Hermitian part of the Hamiltonian.
As in the general case this implies that the Born rule is unstable.

\section{Examples of quantum instability}

The gravitational instability of the Born rule has been studied for several
examples. These include a scalar field on de Sitter space, a scalar field
close to an evaporating black hole, and an atomic system in the gravitational
field of the earth. Here we outline the results obtained so far and some of
the potential implications.\footnote{For more details see ref. \cite{AV21}.}

\subsection{Inflationary perturbations on de Sitter space}

In ref. \cite{AV21}, taking the results of Brizuela et al. \cite{BKM16} as a
starting point, we derived a simplified model for inflationary perturbations
in a far slow-roll limit, on a background with an approximate de Sitter
expansion, $a\propto e^{Ht}$, where the Hubble parameter $H$ is almost
constant. The resulting equations define a tractable cosmological model of
quantum instability in the early inflationary universe.

The perturbations are described by (real) Fourier field components
$q_{\mathbf{k}}$. The corrected Schr\"{o}dinger equation for the effective
wave function $\psi_{\mathbf{k}}^{(1)}(q_{\mathbf{k}},t)$ is found to be%
\begin{equation}
i\frac{\partial\psi_{\mathbf{k}}^{(1)}}{\partial t}=\hat{H}_{\mathbf{k}}%
\psi_{\mathbf{k}}^{(1)}-\frac{\bar{k}^{3}}{2m_{\mathrm{P}}^{2}H^{2}}\frac
{1}{\psi_{\mathbf{k}}^{(0)}}\left[  \frac{1}{a^{3}}(\hat{H}_{\mathbf{k}}%
)^{2}\psi_{\mathbf{k}}^{(0)}+i\frac{\partial}{\partial t}\left(  \frac
{1}{a^{3}}\hat{H}_{\mathbf{k}}\right)  \psi_{\mathbf{k}}^{(0)}\right]
\psi_{\mathbf{k}}^{(1)}\ , \label{Sch_corr_deS}%
\end{equation}
where%
\begin{equation}
\hat{H}_{\mathbf{k}}=-\frac{1}{2a^{3}}\frac{\partial^{2}}{\partial
q_{\mathbf{k}}^{2}}+\frac{1}{2}ak^{2}q_{\mathbf{k}}^{2} \label{H_k}%
\end{equation}
is the uncorrected (zeroth-order) Hamiltonian for the field mode,
$\psi_{\mathbf{k}}^{(0)}$ is the uncorrected (zeroth-order) wave function, and%
\begin{equation}
\bar{k}=\frac{1}{\mathfrak{L}}\ , \label{kbar}%
\end{equation}
where $\mathfrak{L}$ is an arbitrary lengthscale associated with spatial
integration in the classical action (to be interpreted as an infrared cutoff)
\cite{Vent14}. In the same limit the de Broglie guidance equation for
$q_{\mathbf{k}}$ is found to be%
\begin{equation}
\frac{dq_{\mathbf{k}}}{dt}=\frac{1}{a^{3}}\frac{\partial s_{\mathbf{k}}^{(1)}%
}{\partial q_{\mathbf{k}}}\ , \label{deB_qc_restored}%
\end{equation}
where $s_{\mathbf{k}}^{(1)}=\operatorname{Im}\ln\psi_{\mathbf{k}}^{(1)}$ is
the phase of $\psi_{\mathbf{k}}^{(1)}$. This is the standard de Broglie
velocity for Fourier components of a scalar field, with Hamiltonian
(\ref{H_k}), on a classical expanding background \cite{AV07}.

For a theoretical ensemble with the same wave function $\psi_{\mathbf{k}%
}^{(1)}(q_{\mathbf{k}},t)$, the probability density $\rho_{\mathbf{k}}%
^{(1)}(q_{\mathbf{k}},t)$ will evolve by the continuity equation%
\begin{equation}
\frac{\partial\rho_{\mathbf{k}}^{(1)}}{\partial t}+\frac{\partial}{\partial
q_{\mathbf{k}}}\left(  \rho_{\mathbf{k}}^{(1)}\dot{q}_{\mathbf{k}}\right)
=0\ , \label{cont_qc}%
\end{equation}
where $\dot{q}_{\mathbf{k}}$ is the velocity field (\ref{deB_qc_restored}). In
contrast, from (\ref{Sch_corr_deS}) we find that $\left\vert \psi_{\mathbf{k}%
}^{(1)}\right\vert ^{2}$ satisfies%
\begin{equation}
\frac{\partial\left\vert \psi_{\mathbf{k}}^{(1)}\right\vert ^{2}}{\partial
t}+\frac{\partial}{\partial q_{\mathbf{k}}}\left(  \left\vert \psi
_{\mathbf{k}}^{(1)}\right\vert ^{2}\dot{q}_{\mathbf{k}}\right)  =s\ ,
\end{equation}
where in the notation of Section 8 the `source' $s$ is given by (\ref{s})
where here%
\begin{equation}
\hat{H}_{2}=-\frac{\bar{k}^{3}}{2m_{\mathrm{P}}^{2}H^{2}}\frac{1}%
{\psi_{\mathbf{k}}^{(0)}}\frac{\partial}{\partial t}\left(  \frac{1}{a^{3}%
}\hat{H}_{\mathbf{k}}\right)  \psi_{\mathbf{k}}^{(0)}\ .
\end{equation}

These equations can be used to calculate the gravitational production of
quantum nonequilibrium during inflation, employing the differential equation
(\ref{dfdt}) for the rate of change of the ratio $f_{\mathbf{k}}%
=\rho_{\mathbf{k}}/\left\vert \psi_{\mathbf{k}}\right\vert ^{2}$ along
trajectories. Taking $\psi_{\mathbf{k}}^{(0)}$ to be the Bunch-Davies vacuum
wave function, approximate calculations show that the gravitational
instability of the Born rule generates a nonequilibrium deficit $\sim1/k^{3}$
in the primordial cosmological power spectrum. It has been shown elsewhere
that there is no significant relaxation during inflation \cite{AV10,KV19}, so
the condition (\ref{condn}) will be satisfied and the generated nonequilibrium
will persist over time. However, the magnitude of the effect on the power
spectrum is far too small to observe in the CMB (for details see ref.
\cite{AV21}).

By considering only the Hermitian terms in the Hamiltonian, Brizuela et al.
\cite{BKM16} show that the gravitationally-corrected wave function induces a
similar $\sim1/k^{3}$ correction to the power spectrum but of opposite sign
(hence a power excess). However, the calculations of ref. \cite{AV21} are too
approximate to precisely compare the overall magnitudes of these
physically-distinct effects.

\subsection{Evaporating black holes}

It is also of interest to consider quantum instability for a field in the
background spacetime of an evaporating Schwarzchild black hole. It was argued
by Kiefer, M\"{u}ller and Singh \cite{KMS94} that in this case the
quantum-gravitational corrections to the effective Schr\"{o}dinger equation
will be as in equations (\ref{Sch_correcns})--(\ref{Sch_correcns_b}) but with
the replacement%
\begin{equation}
\sqrt{g}R\rightarrow-16\pi GM/c^{2}\ ,\label{rep_BH}%
\end{equation}
where the Schwarzchild radius $r_{\mathrm{S}}=2GM/c^{2}$ (for a black hole of
mass $M$) provides a natural lengthscale. The non-Hermitian term in
(\ref{Sch_correcns}) reads%
\begin{equation}
i\hat{H}_{b}=i\frac{4\pi\hslash G}{c^{4}}\int d^{3}x\ N\frac{\delta}%
{\delta\tau}\left(  \frac{\mathcal{\hat{H}}_{\phi}}{\sqrt{g}R}\right)
\ ,\label{non-H_dt}%
\end{equation}
where we have inserted $\mu=c^{2}/32\pi G$ as well as $\hbar$ and $c$. With
the replacement (\ref{rep_BH}), (\ref{non-H_dt}) takes the approximate form%
\begin{equation}
i\hat{H}_{b}\simeq-i\frac{\hslash}{4c^{2}}\frac{d}{dt}\left(  \frac{1}%
{M}\right)  \int d^{3}x\ N\mathcal{\hat{H}}_{\phi}=i\frac{\hslash}{4c^{2}%
}\frac{1}{M^{2}}\frac{dM}{dt}\hat{H}_{\phi}\ ,\label{KS_BH}%
\end{equation}
where $\hat{H}_{\phi}$ is the uncorrected field Hamiltonian (neglecting the
rate of change of $\mathcal{\hat{H}}_{\phi}$ compared with the rate of change
of the background geometry). Kiefer et al. suggested that this term might
alleviate the problem of black-hole information loss (though such a term is
inconsistent with the standard quantum formalism).

Taking the phenomenological time dependence \cite{DeW75,Wald84}%
\begin{equation}
M(t)\simeq M_{0}\left(  1-\kappa\frac{m_{\mathrm{P}}^{3}}{M_{0}^{3}}\left(
\frac{t}{t_{\mathrm{P}}}\right)  \right)  ^{1/3}\ , \label{M(t)}%
\end{equation}
with $M_{0}$ the initial mass, $\kappa$ a numerical factor, and here
$m_{\mathrm{P}}=\sqrt{\hbar c/G}\simeq10^{-5}\ \mathrm{g}$ the standard Planck
mass, we have $dM/dt\simeq-\frac{1}{3}\kappa(m_{\mathrm{P}}/t_{\mathrm{P}%
})\left(  m_{\mathrm{P}}/M\right)  ^{2}$. According to (\ref{KS_BH}) the
Hamiltonian $\hat{H}_{\mathbf{k}}$ of a field mode then acquires a
non-Hermitian correction $i\hat{H}_{2}$ (in the notation of Section 8) with%
\begin{equation}
\hat{H}_{2}\simeq-\frac{1}{12}\kappa\left(  \frac{m_{\mathrm{P}}}{M}\right)
^{4}\hat{H}_{\mathbf{k}}\ . \label{H2_BH}%
\end{equation}
This correction is significant in the final stage of evaporation when $M$
approaches $m_{\mathrm{P}}$, suggesting that the final burst of Hawking
radiation could contain significant departures from the Born
rule.\footnote{The creation of quantum nonequilibrium by evaporating black
holes was previously suggested as a possible mechanism for resolving the
information-loss puzzle \cite{AV04a,AV07,AV14,KV20}, but without a clear
theoretical underpinning.}

Quantum nonequilibrium is expected to be created on a timescale $\tau
_{\mathrm{noneq}}$ of order (\ref{tau_est}), which depends inversely on the
equilibrium mean energy $E_{\mathbf{k}}=\left\langle \hat{H}_{\mathbf{k}%
}\right\rangle $. If we take $E_{\mathbf{k}}\sim k_{\mathrm{B}}T_{\mathrm{H}}$
where%
\begin{equation}
k_{\mathrm{B}}T_{\mathrm{H}}=\frac{\hbar c^{3}}{G}\frac{1}{8\pi M}=\frac
{1}{8\pi}m_{\mathrm{P}}c^{2}\frac{m_{\mathrm{P}}}{M}%
\end{equation}
is the Hawking temperature, then from (\ref{tau_est}) and (\ref{H2_BH}) we
have%
\begin{equation}
\tau_{\mathrm{noneq}}\sim\frac{48\pi}{\kappa}t_{\mathrm{P}}\left(  \frac
{M}{m_{\mathrm{P}}}\right)  ^{5}\ . \label{tau_BH}%
\end{equation}
Corrections to the Born rule will be significant if $\tau_{\mathrm{noneq}}$ is
not too large compared to the evaporation timescale $t_{\mathrm{evap}}$.
Taking $1/t_{\mathrm{evap}}\sim(1/M)\left\vert dM/dt\right\vert $ we have%
\begin{equation}
t_{\mathrm{evap}}\sim\frac{3}{\kappa}t_{\mathrm{P}}\left(  \frac
{M}{m_{\mathrm{P}}}\right)  ^{3} \label{t_evap}%
\end{equation}
and a ratio%
\begin{equation}
\frac{\tau_{\mathrm{noneq}}}{t_{\mathrm{evap}}}\sim\frac{48\pi}{3}\left(
\frac{M}{m_{\mathrm{P}}}\right)  ^{2}%
\end{equation}
(the factor $\kappa$ cancels). Again it seems clear that significant
deviations from the Born rule can be generated in the outgoing radiation only
in the final stage of evaporation when $M$ approaches $m_{\mathrm{P}}$.

It is however not known if such deviations could survive quantum relaxation,
which may well be significant in the final stage of evaporation when the
background spacetime is changing rapidly. Quantum nonequilibrium will build up
over time only if (\ref{condn}) is satisfied in the regime where $M$
approaches $m_{\mathrm{P}}$. Thus we need to know how $\tau_{\mathrm{relax}}$
scales with $M$ and to compare this with our estimate $\tau_{\mathrm{noneq}%
}\propto\left(  M/m_{\mathrm{P}}\right)  ^{5}$. This is a matter for future work.

Should nonequilibrium survive in the outgoing radiation, at least in principle
the emitted photons could show anomalies in their two-slit interference
pattern or in their polarisation probabilities \cite{AV04b}. Realistically,
Hawking radiation in the $\gamma$-ray region might be detected from exploding
primordial black holes (which may form a significant component of dark matter
\cite{primBHs}). However, only the very final burst is likely to show
significant deviations from the Born rule, making detection difficult.

\subsection{An atom in the gravitational field of the earth}

We might ask if the Born rule could be unstable for an atomic system in the
gravitational field of the earth. We saw in Section 10.2 that the
non-Hermitian correction to a field Hamiltonian in the spacetime of a
Schwarzchild black hole can plausibly be obtained from (\ref{non-H_dt}) by
replacing $\sqrt{g}R$ by $-8\pi r_{\mathrm{S}}$ where $r_{\mathrm{S}%
}=2GM/c^{2}$ is the natural lengthscale of the background. In the
gravitational field of the earth we might expect instead to make a replacement
of the form%
\begin{equation}
\sqrt{g}R\rightarrow-8\pi r_{\mathrm{c}}\ , \label{lab}%
\end{equation}
where $r_{\mathrm{c}}$ is the local radius of curvature ($r_{\mathrm{c}}%
\simeq10^{13}\ \mathrm{cm}$ at the surface of the earth).\footnote{A similar
suggestion was made by Kiefer and Singh \cite{KS91}, who considered the effect
of the Hermitian correction on atomic energy levels.} Inserting this in
(\ref{non-H_dt}), and writing $\mathcal{\hat{H}}_{\phi}$ as $\mathcal{\hat{H}%
}_{a}$ where $\hat{H}_{a}=\int d^{3}x\,N\mathcal{\hat{H}}_{a}$ is the atomic
Hamiltonian, we find an estimated non-Hermitian term%
\begin{equation}
i\hat{H}_{b}\sim-i\frac{\hslash G}{c^{4}}\frac{1}{r_{\mathrm{c}}}\int
d^{3}x\ N\frac{\delta}{\delta\tau}\left(  \mathcal{\hat{H}}_{a}\right)
\sim-i\frac{l_{\mathrm{P}}}{r_{\mathrm{c}}}\frac{\partial\hat{H}_{a}}{\partial
t}t_{\mathrm{P}}\ , \label{Ham_corr_atomic}%
\end{equation}
where $l_{\mathrm{P}}$ and $t_{\mathrm{P}}$ are respectively the Planck length
and time.

The term (\ref{Ham_corr_atomic}) is non-zero only if the (uncorrected) atomic
Hamiltonian $\hat{H}_{a}$ is time dependent. The magnitude of
(\ref{Ham_corr_atomic}) is roughly the change in $\hat{H}_{a}$ over a Planck
time suppressed by the ratio $l_{\mathrm{P}}/r_{\mathrm{c}}$. Needless to say,
in ordinary laboratory conditions, this term will be utterly negligible.
Furthermore, if $\hat{H}_{a}$ changes rapidly (to maximise the effect), the
atomic wave function will be a superposition of multiple energy eigenstates,
and we expect to find quantum relaxation over timescales $\tau_{\mathrm{relax}%
}<<\tau_{\mathrm{noneq}}$. Even if we could probe an atomic ensemble over
times $\sim\tau_{\mathrm{noneq}}$ (far longer than the age of the universe),
any gravitationally-generated nonequilibrium will have long-since relaxed. It
then appears that the gravitational creation of quantum nonequilibrium in
ordinary laboratory systems -- with a dynamical Hamiltonian in a background
curved space -- is likely to be of theoretical interest only.

\section{Conclusion}

We have argued that, in the deep quantum-gravity regime, with a
non-normalisable Wheeler-DeWitt wave functional $\Psi$, there is no Born rule
and the universe is in a perpetual state of quantum nonequilibrium with a
probability density $P\neq\left\vert \Psi\right\vert ^{2}$. Quantum relaxation
to the Born rule can occur only when the early universe emerges into a
semiclassical or Schr\"{o}dinger approximation, with a time-dependent and
normalisable effective wave functional $\psi$ for a system on a classical
spacetime background, for which the probability density $\rho$ can evolve
towards $\left\vert \psi\right\vert ^{2}$ (on a coarse-grained level). We
conclude that the long-standing hypothesis of primordial quantum
nonequilibrium, with relaxation to the Born rule taking place soon after the
big bang, follows naturally from the internal logic of quantum gravity (as
interpreted in de Broglie-Bohm pilot-wave theory). Furthermore,
quantum-gravitational corrections to the Schr\"{o}dinger approximation, in the
form of tiny non-Hermitian terms in the effective Hamiltonian, generate a
(very slight) instability of the Born rule, whereby quantum nonequilibrium
$\rho\neq\left\vert \psi\right\vert ^{2}$ can be created from a prior
equilibrium ($\rho=\left\vert \psi\right\vert ^{2}$) state. To observe such
effects will be difficult in practice, though possible at least in principle.

When restricted to the Born-rule equilibrium state, the pilot-wave or de
Broglie-Bohm formulation of quantum theory is experimentally indistinguishable
from textbook quantum mechanics. Wider support for this formulation is likely
to be forthcoming should we find experimental evidence for violations of the
Born rule -- or if the theory allows us to make decisive progress in
understanding some vital aspect of fundamental physics. From the results
presented here we suggest that this little-used formulation of quantum theory
allows us to understand and solve three problems in canonical quantum gravity:
(a) to explain why the naive Schr\"{o}dinger interpretation does not work, (b)
to account for the emergence of the Born rule in a semiclassical regime, and
(c) to give a consistent meaning to non-Hermitian quantum-gravitational
corrections to the effective Schr\"{o}dinger equation.

The results of this paper also impact on certain philosophical debates
concerning the status of the Born rule in de Broglie-Bohm theory. As we have
noted, and discussed in detail elsewhere \cite{Allori20}, the `Bohmian
mechanics' school employs an essentially circular argument to obtain the Born
rule for subsystems by assuming the Born rule for the whole universe at some
initial cosmological time.\footnote{In a remarkable reply to ref.
\cite{Allori20}, D\"{u}rr and Struyve \cite{DS21} invoke similar circular
reasoning in their account of classical coin tossing.} We have seen that, when
quantum gravity is taken into account, such an argument has no starting point,
since there is no fundamental Born-rule measure for a universe governed by the
Wheeler-DeWitt equation (despite attempts by some workers \cite{Vink92,DS20}
to apply the naive Schr\"{o}dinger interpretation to pilot-wave gravitation).

It is one hundred years since de Broglie started on the path that, after five
years of remarkable developments, brought him in 1927 to pilot-wave theory as
we know it today. It is seventy years since the revival and further
development of pilot-wave theory in Bohm's papers of 1952. And yet the theory
is still not widely known or used, and is often misunderstood. The historical
development of pilot-wave theory is in certain respects reminiscent of the
historical development of the kinetic theory of gases. Beginning with the
pioneering work of Bernoulli in the early 18th century, and of Herapath and
Waterston in the early 19th century, kinetic theory was more or less ignored
until it was taken up by Clausius in an influential paper of 1857 \cite{SB61}.
Another half a century had to pass, with decisive contributions in particular
by Maxwell, Boltzmann and Einstein, before theorists were able to interpret
Brownian motion as evidence for atoms and kinetic theory. Whether or not a
comparable empirical breakthough awaits pilot-wave theory remains to be seen.

Why were physicists in the late nineteenth-century still reluctant to accept
the existence of atoms and molecules, long after chemists had already deduced
their detailed shapes and compositions? In part there was philosophical
opposition from Mach and others, who emphasised the role of sensory perception
in physics, while the idea of an objective reality beyond the immediate reach
of our senses came to be widely derided as unscientific and metaphysical.
Similarly, today there remains widespread opposition to realism in quantum
physics. For as long as the details of de Broglie-Bohm trajectories cannot be
observed (the uncertainty principle reigns for as long as we are confined to
quantum equilibrium) those trajectories will continue to be dismissed as unphysical.

A decisive breakthrough, with an end to seemingly endless philosophical
debates, will occur only by extending the boundaries of physics beyond what is
currently known and understood. The prospects do not seem entirely remote. As
we have argued in this paper, gravitation may hold the key to unlocking the
hidden physics of pilot-wave theory.

\end{document}